\documentclass{aa}  

\usepackage{natbib}

\usepackage{caption}
\usepackage{subcaption}

\usepackage[flushleft]{threeparttable}

\usepackage{graphicx}
\usepackage{txfonts}
\usepackage{gensymb}

\usepackage{url}

\makeatletter
\newcommand\footnoteref[1]{\protected@xdef\@thefnmark{\ref{#1}}\@footnotemark}
\makeatother

\begin{document}

   \title{The faint radio sky: VLBA observations of the COSMOS field\thanks{Table~\ref{table:agncat} and Table~\ref{table:mw} are only available in electronic form at the CDS via anonymous ftp to cdsarc.u-strasbg.fr (130.79.128.5) or via http://cdsweb.u-strasbg.fr/cgi-bin/qcat?J/A+A/}}


   \author{N. Herrera Ruiz
          \inst{1}
          \and
          E. Middelberg\inst{1}
          \and
          A. Deller\inst{2,3}
          \and
          R. P. Norris\inst{4,5}
          \and
          P. N. Best\inst{6}
          \and
          W. Brisken\inst{7}
          \and
          E. Schinnerer\inst{8}
          \and
          V. Smol{\v c}i{\'c}\inst{9}
          \and
          I. Delvecchio\inst{9}
          \and
          E. Momjian\inst{7}
          \and
          D. Bomans\inst{1}
          \and
          N. Z. Scoville\inst{10}
          \and
          C. Carilli\inst{7}
          }

   \institute{Astronomisches Institut, Ruhr-Universit\"at Bochum, 
              Universit\"atstrasse 150, 44801 Bochum, Germany\\
              \email{herrera@astro.rub.de}
         \and The Netherlands Institute for Radio Astronomy (ASTRON), Dwingeloo, The Netherlands
             \and Centre for Astrophysics and Supercomputing, Swinburne University of Technology, P.O. Box 218, Hawthorn, VIC 3122, Australia
             \and CSIRO Australia Telescope National Facility, PO Box 76, Epping, NSW 1710, Australia
             \and Western Sydney University, Locked Bag 1797, Penrith South, NSW 1797, Australia
             \and SUPA, Institute for Astronomy, Royal Observatory Edinburgh, Blackford Hill, Edinburgh EH9 3HJ, UK
             \and National Radio Astronomy Observatory, P.O. Box O, Socorro, NM 87801, USA
             \and MPI for Astronomy, K\"onigstuhl 17, D-69117 Heidelberg, Germany
             \and Department of Physics, Faculty of Science, University of Zagreb, Bijeni\v{c}ka cesta 32, 10000 Zagreb, Croatia
             \and California Institute of Technology, MC 249-17, 1200 East California Boulevard, Pasadena, CA 91125, USA
             }

   \date{Received May 12, 2017; accepted July 13, 2017} 

 
  \abstract
      {Quantifying  the fraction of active galactic nuclei in the faint radio population and understanding their relation with star-forming activity are fundamental to studies of galaxy evolution. Very long baseline interferometry (VLBI) observations are able to identify active galactic nuclei (AGN) above relatively low redshifts (z > 0.1) since they provide milli-arcsecond  resolution. 
      }
   {We have created  an AGN catalogue from 2865 known radio sources observed in the Cosmic Evolution Survey (COSMOS) field, which has   exceptional multi-wavelength coverage. With this catalogue we intend to study the faint radio sky with statistically relevant numbers and to analyse the AGN -- host galaxy co-evolution, making use of the large amount of ancillary data available in the field. }
   {Wide-field VLBI observations were made of all known radio sources in the COSMOS field  at 1.4 GHz to measure the  AGN fraction, in particular in the faint radio population. We describe in detail the observations, data calibration, source detection and flux density measurements, parts of which we have developed for this survey. The combination of number of sources, sensitivity, and area covered  with this project are unprecedented.}
   {We have detected 468 radio sources, expected to be AGNs, with the Very Long Baseline Array (VLBA). This is, to date, the largest sample assembled of VLBI detected sources in the sub-mJy regime. The input sample was taken from  previous observations with the Very Large Array (VLA). We present the catalogue with additional optical, infrared and X-ray information.}
   {We find a detection fraction of 20$\pm$1\%, considering only those sources from the input catalogue which were in principle detectable with the VLBA (2361). As a function of the VLA flux density, the detection fraction is higher for higher flux densities, since at high flux densities a source could be detected even if the VLBI core accounts for a small percentage of the total flux density. As a function of redshift, we see no evolution of the detection fraction over the redshift range $0.5<z<3$. In addition, we find that faint radio sources typically have a greater fraction of their radio luminosity in a compact core -- $\sim$70\% of the sub-mJy sources detected with the VLBA have more than half of their total radio luminosity in a VLBI-scale component, whereas this is true for only $\sim$30\% of the sources that are brighter than 10\,mJy.  This suggests that fainter radio sources differ intrinsically from brighter ones. Across our entire sample, we find the predominant morphological classification of the host galaxies of the VLBA detected sources to be early type (57\%), although this varies with redshift and at z>1.5 we find that spiral galaxies become the most prevalent (48\%). The number of detections is high enough to study the faint radio population with  statistically significant numbers. We demonstrate that wide-field VLBI observations,  together with new  calibration methods such as multi-source self-calibration and mosaicing, result in information which is difficult or impossible to obtain otherwise. }


   \maketitle
%

\section{Introduction}
\label{sec:int}

The main motivation of studying faint radio sources is to understand
how active galactic nuclei (AGN) and star formation evolve through
cosmic time. In particular, AGN appear to be fundamental players in galaxy evolution and star formation, which makes it necessary to determine where an AGN
is present. Radio surveys are indispensable components of large
multiwavelength studies since they are not affected by dust and they can detect the non-thermal radiation from AGNs. 

A strongly debated topic in astrophysics related to AGN and star formation interplay is the suggested link between accretion activity in AGN and star-forming activity of the host galaxy by AGN `feedback' (e.g. \citealt{best2006, croton2006}). In recent studies, two different accretion modes have been put forward for AGN-host galaxy co-evolution (e.g. \citealt{best2014}): the cold-mode (radiatively efficient) and the hot-mode (radiatively inefficient). The cold-mode AGNs are typically associated with star-forming galaxies and low mass black holes fuelled by cold gas of a thin, optically thick accretion disk with a high accretion rate (e.g. \citealt{heckman2014, norris2012, smolcic2009}). The hot-mode AGNs are primarily associated with massive black holes hosted by elliptical galaxies and likely fuelled by hot gas of the halo leading to a slow growth of the black hole with a low accretion rate and a low or nonexistent star formation (e.g. \citealt{bower2006, schawinski2009, dubois2013}). The main result of this energetic process is the outflow of collimated jets. For a more detailed description of these two modes, see \citet{heckman2014}. Although this division of the two modes seems to work fine at low redshifts, it might not be right at high redshifts. \citet{rees2016} investigated the host galaxy properties of a sample of radio-loud AGN and found that the majority of z\,>\,1.5 radio-AGN are hosted by star-forming galaxies. \citet{zinn2013} suggested that both cold- and hot-mode mechanisms are important and showed that the star formation rate is correlated with radio jet power. Nevertheless, the specific astrophysics behind this relation are still not well understood.

\citet{strazzullo2010} analysed the rest-frame U – B versus B colour-magnitude diagram of their radio selected sample and found that most of their sources were located in an intermediate location and not in two different locations, suggesting that at faint flux densities a simple classification between AGNs or star-forming galaxies might not be appropriate. The properties of this intermediate location are not well understood yet, and it is composed of a mixed population of AGN, star-forming and composite galaxies where star formation and activity from the nucleus both play an important role. In addition, it has been shown in several cases that galaxies with a spectral energy distribution (SED) typical of a star-forming galaxy actually have the radio luminosity or morphology of an AGN. \citet{norris2009}
suggest that such sources represent a class of AGN buried deeply inside a dusty star-forming galaxy, appearing to be an increasingly common phenomenon at high redshifts (z$\gtrsim$1).

It is therefore of considerable interest to produce AGN catalogues. However, it is usually difficult or impossible to distinguish between AGN and star-forming galaxies because radio observations are typically carried out with
interferometers such as the Jansky Very Large Array or the Australia
Telescope Compact Array. Because of their limited baseline lengths of
several km or several tens of km, these instruments are equally
sensitive to radio emission from either process. Fortunately, a
relatively direct way to identify which galaxies do have radio-active
AGN is a detection with very long baseline interferometry (VLBI)
observations. VLBI baselines are typically a few thousand km long,
resulting in an angular resolution of the order of milli-arcsec. The brightness temperature a body needs to have to be detected using VLBI is around $10^{6}$\,K, which generally can only be reached by the non-thermal emission processes in AGN \citep{condon1992}. At a redshift of 0.1, the actual diameter of an object that can be resolved with the Very Long Baseline Array (VLBA) is 10\,pc. \citet{kewley2000} show that in compact objects such brightness temperatures can only be achieved by AGN activity which makes this technique a powerful method to cleanly separate AGNs from star-forming galaxies.

The main disadvantage of VLBI is its tiny field of view, which covers a radius of only around 5\,arcseconds at GHz frequencies. Consequently, this technique has historically been incompatible with large field observations, targeting a significant number of sources. However, a new technique has been developed to deal with this problem, the so-called ``wide-field VLBI'' technique \citep{garrett1999}, whose main objective is to make the field of view in VLBI observations as wide as possible. \citet{garrett2001} presented deep, wide-field VLBI observations at 1.6\,\,GHz of the Hubble deep field (HDF) region. They detected two radio sources at the 5$\sigma$ level and a third radio source was detected at the 4$\sigma$ level. \citet{garrett2005} conducted deep, wide-field VLBI observations at 1.4\,GHz of an area of the sky located within the NOAO Bootes field. They observed 61 sources and detected nine sources above the 6$\sigma$ level. \citet{lenc2008} performed the first wide-field VLBI survey at 90\,cm. They targeted 618 sources in an area consisting of two overlapping fields centred on the quasar J0226+3421 and the gravitational lens B0218+357 and detected 27 sources out of the 272 detectable sources. \citet{chi2013} carried out wide-field VLBI observations of the Hubble deep field north (HDF-N) and flanking fields (HFF). They observed 92 known radio sources with a global VLBI array at 1.4\,GHz and detected 12 sources above the 5$\sigma$ level. The development of wide-field VLBI was limited by the spectral and temporal resolution of the early generation of hardware correlators. To make this technique feasible, important progress in computer technology was required and the introduction of software correlators played a decisive role \citep{deller2007, deller2011}.

With the improvement in sensitivity of radio interferometers, the minimum of brightness temperature needed for a source to be detected can occasionally be reached by star-forming activity, radio supernovae or gamma-ray bursts. However, the luminosity of star formation quickly drops below the detection threshold when the galaxies are located beyond a redshift of 0.1, where almost all of our targets are located. Furthermore, the transient events are exceedingly rare: after $\sim$30 years of observations only $\sim$50 supernovae have been detected at radio wavelengths, none of which are Type Ia (the most powerful ones) \citep{lien2011}, and from a sample of 304 Gamma-Ray Burst observed with radio telescopes (during 14 years) the fractional detection rate of radio afterglows is about 30\% \citep{chandra2012}. Therefore, we are confident that our sample of VLBI detected radio sources constitutes a pure sample of radio-active AGN.

We have observed 2865 known radio sources from \citet{schinnerer2010} with the VLBA in the The Cosmic Evolution Survey (COSMOS) field. COSMOS is an astronomical survey designed to probe the formation and evolution of galaxies as a function of cosmic time and large-scale structural environment \citep{scoville2007}. The COSMOS field is located at \mbox{RA (J2000) = 10:00:28.6} and \mbox{DEC (J2000) = +02:12:21.0} and is suitable to study the radio AGN-host galaxy interplay, since it is certainly the most comprehensive extragalactic survey to date. COSMOS includes very sensitive radio, sub-mm, infrared, optical and X-ray data from diverse facilities\footnote{\url{http://cosmos.astro.caltech.edu/page/datasets}} and provides a unique multi-wavelength coverage over an area as large as 2\,deg$^{2}$. Moreover, it is ideally suited to study the faint radio sky, since the COSMOS field is mostly lacking even moderately strong radio sources.

In this paper, we describe the VLBA observations and we present the resulting AGN catalogue. We made use of several specialised wide-field VLBI techniques such as mosaicing, multi-source self-calibration, and primary beam corrections, to generate milli-arcsecond scale resolution images with a sensitivity of tens of $\mu$Jy. We have previously used wide-field VLBI observations to carry out similar observations for a wide range of flux densities \citep{middelberg2011, middelberg2013, deller2014}, demonstrating the feasibility of the process.


The structure of this paper is as follows. In Section~\ref{sec:cal} we describe the details of the observations and the procedure of the data calibration. We present the catalogue of the VLBA detected sources in Section~\ref{sec:cat}. In Section~\ref{sec:res} the results of the observations are reported and discussed. In Section~\ref{sec:con} we summarise the conclusions derived by the present project.

Throughout this paper, we assume a flat $\Lambda$CDM cosmology with
${\it{H}}_{0}$ = 67.3 $km$ $s^{-1}$ $Mpc^{-1}$, $\Omega_{M}$ = 0.31
and $\Omega_{\Lambda}$ = 0.69 (according to the recent Planck results
published by \citealt{planck2014}).


\section{Observations and data calibration}
\label{sec:cal}

In this section, we describe our sample, the characteristics of the observations and the procedures followed to calibrate the data. We also present the obtained sensitivity map and the description of the criteria adopted for source extraction. For clarity, we establish the nomenclature used here:

The term `target' refers to an object to be observed with the VLBA. 

The term `source' refers to a physical source.

The term `component' refers to a connected region of radio emission.

The term `pointing' refers to one of the 23 regions, within which objects were targeted during an observing run.

The term `epoch' refers to six hours observation of one pointing.

The term `phase centre' refers to a location within a telescope's beam where correlation can focus. 

The term `calibration file' refers to the data set of each epoch, containing calibration information from the phase-referencing source J1011+0106 and the fringe-finder 4C\,39.25.










\subsection{Sample}

We took the target positions for our sample from the VLA catalogue of \citet{schinnerer2010}. Because the individual fields of view in VLBI observations are smaller than the extension of the source as seen on arcsec scales, we decided to target each component of the sources classified as multi-component by the initial catalogue. In particular, from the 2865 radio sources, 131 were multi-component sources, yielding a total of 3293 targets.

\subsection{Observations}

\begin{figure}
\centering
\includegraphics[scale=0.37]{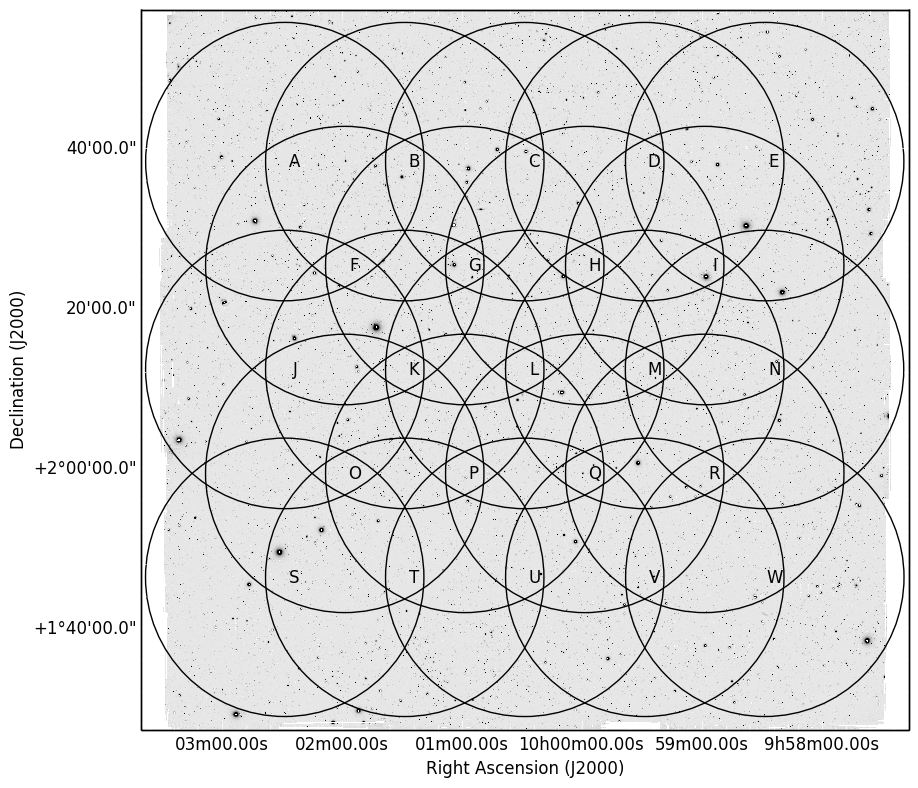}
\caption{\em {The 23 pointings representing the design of our VLBA
    observations of the COSMOS field. The black circles denote the
    radius within which sources were targeted during an observing
    run ($\sim$15\arcmin). The letters denote the identification for each pointing. The
    background greyscale image is a mosaic of COSMOS Subaru i-band
    data\protect\footnotemark.}}
\label{fig:design}
\end{figure}

We observed 3293 targets in the COSMOS extragalactic field
with the VLBA at a central frequency of 1.54 GHz over 23 pointings between February 2012 and January 2013. To compensate for the reduced sensitivity at the edge of the pointings, we used a pattern of overlapping pointings (radius $\sim$15\arcmin), which is
standard practice with compact interferometers (see
Fig.~\ref{fig:design}). For this reason, most of the targets were
observed several times. Each pointing was observed twice for 6
hours, to increase observing time at high elevations and to allow for
greater scheduling flexibility. This resulted in 46 individual
observing epochs. During each epoch, the target field was repeatedly observed for 4.5 min, followed by a
1 min observation of the phase-referencing source J1011+0106. For data
consistency checks, the fringe-finder 4C\,39.25 was observed every 2\,h. In summary, the on-source integration time per pointing was roughly 8.5 hours, with the maximum number of overlapping pointings being seven. Eight
32-MHz bands were observed in two circular polarisations, requiring a
recording rate of 2Gbps. A minimum number of nine VLBA antennas was
scheduled to achieve the required sensitivity.

\footnotetext{\url{http://irsa.ipac.caltech.edu/data/COSMOS/}}

Historically, VLBI observations have provided the highest resolution
in astronomy with the drawback of covering only tiny fields of view
(around 5\,arcseconds radius at GHz frequencies) as a result of
  the high fringe rates implied in VLBI observations. This has made them unsuitable for observing
large fields with a considerable number of objects. However, a new
multi-phase centre mode has been developed for the VLBA DiFX
correlator in operation at the VLBA \citep{deller2007}. In this mode, the initial
  correlation is carried out with high time and frequency
  resolution. The visibilities are subsequently phase-shifted to the
  other phase centres, before they are averaged in time and frequency
  and written to disk. This minimises the effects of time and
  bandwidth averaging, while at the same time the resulting data sets
  are kept comparatively small. Thus, within a region confined by the
  individual antenna's primary beams one can position numerous phase
  centres. This mode can be used to
image hundreds of objects in a single observing run
\citep{deller2011}. In our case, an average of 450 VLA sources per
pointing were targeted. The raw dataset size of each individual target
was 350\,MB, yielding around 158\,GB per epoch on average. Since we had 46 epochs, the total amount of raw data
from our observations was $\sim$7 TB, indicating that processing was a
significant computing effort.

\subsection{Data calibration}
\label{sec:calib}

We have calibrated the data using the Astronomical Image Processing
System (\texttt{AIPS}, \citealt{greisen2003, fomalont1981}) following standard
procedures used in phase-referenced VLBI observations together
  with specialised techniques developed for wide-field VLBI
  observations. Our script to calibrate the data has been written in
\texttt{ParselTongue} (\texttt{AIPS} Talking \texttt{Python},
\citealt{kettenis2006}). The details of our calibration procedure are described
as follows. The specialised steps of wide-field VLBI are refered to as `non-standard calibration steps':

\subsubsection{Loading and initial preparation of the uv data}

We loaded the data into AIPS using the task FITLD. If there were redundancies in prior calibration information, such as system temperature and gain curves, on the same source and/or antenna in a given data set, we used the procedure MERGECAL to remove the redundant information. We sorted the visibility data set into time-baseline order using the task UVSRT.

This is a non-standard calibration step. Each epoch consists of one calibration file and data sets containing measurements of each target. First, we use the calibration file to find the corrections to calibrate the data. 

\subsubsection{First corrections}

Since the parallactic angle between the calibrator and target is
  different at different stations, we corrected for this phase term
  using the AIPS task CLCOR.

The Earth orientation parameters (EOPs) used at correlation time are later
  updated and refined, resulting in a change of observed visibility
  phase. To correct for this, we obtained updated EOPs from the USNO
  server\footnote{\url{http://gemini.gsfc.nasa.gov/solve_save/usno_finals.erp}}
  and applied corrections using the task CLCOR.

The ionosphere can cause unmodelled dispersive delays that we
  corrected using
  measurements\footnote{\url{ftp://cddis.gsfc.nasa.gov/gps/products/ionex}}
  of total electron content (TEC) and the task TECOR.

We corrected the amplitude offsets arising from sampler
  errors at the stations, typically of order 5-10\%, using the task ACCOR.

\subsubsection{Phase, bandpass and amplitude calibration}

Residual delays, rates and phases were measured using data from
  the phase calibrator J1011+0106 and the fringe-finder 4C 39.25,
  using the task FRING. We used a solution interval of two minutes and
  averaged the data in each IF of 32\,MHz.

A recent VLBA Scientific
  Memo\footnote{\url{http://library.nrao.edu/public/memos/vlba/sci/VLBAS_37.pdf}}
  reports that the then standard amplitude calibration of the VLBA
  caused amplitude errors of order 25\% to 30\%. The memo
    recommends to form a model bandpass using the full band and power
    normalisation and to scale the data by a small factor to make the
    calibrated autocorrelation values unity. We implemented this
    procedure using the task BPASS on the fringe-finder, and the new
    AIPS task ACSCL to deal with the small offset from unity amplitude
    in the calibrated autocorrelations.

We carried out amplitude calibration using the antennas' system temperature ($T_{\rm sys}$)
    measurements and known gain curves with the task APCAL. $T_{\rm sys}$ depends on
  the antenna zenith angles, $\Theta_{z}$, and follows approximately
  1/cos($\Theta_{z}$). Because of the low declination of the field
  (2\degree) and the limited window of six hours observing time, the
  zenith angle (and therefore antenna elevation) at most stations was
  roughly constant throughout the observations and the $T_{\rm sys}$
  value were expected to vary only little. Typical $T_{\rm sys}$ were
  found to be of order 30\,K, and values exceeding 50\,K were deemed
  to indicate errors. These values were flagged, which resulted in the
  respective visibility data not being used in subsequent processing
  steps. We consider that the surface brightness (SB) error resulting from calibration is of the order of 10\% \citep{middelberg2013}. We used this error later to calculate the errors of the VLBA flux densities (see Section~\ref{subsec:flxpos}).

\subsubsection{Flagging}

We edited the data using a flagging programme written by us
  and implementing the procedures described in \citet{middelberg2006},
  which compare the median amplitude in each channel to the median of
  an RFI-free reference channel. Around 4\%-5\% of the data were flagged.

\subsubsection{Multi-source self-calibration}

\begin{figure*}
\centering
\includegraphics[scale=0.6]{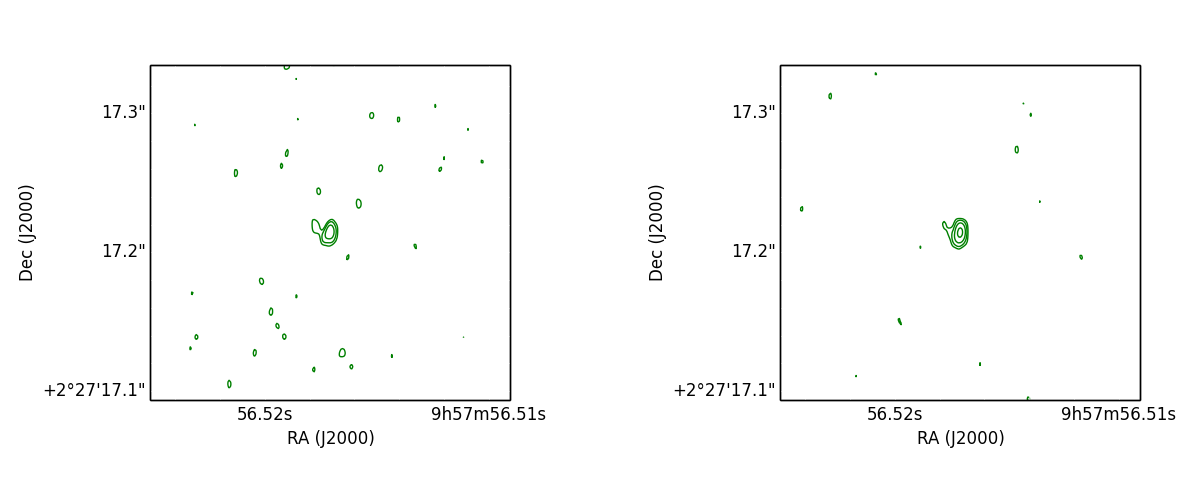}
\caption{\em {Contour plots of a target before applying
    multi-source self-calibration (left) and after applying
    multi-source self-calibration (right). The peak flux density of the left panel image is 1.2\,mJy and the rms is 58\,$\mu$Jy. The peak flux density of the right panel image is 1.6\,mJy and the rms is 55\,$\mu$Jy. Positive contours start at
    three times the rms level of the image and increase by a factor of
    two. The image is uniformly-weighted. The peak flux density has
    increased and the sidelobe level is reduced.}}
\label{fig:multifield}
\end{figure*}

This is a non-standard calibration step. In phase-referenced observations, images have reduced coherence due to ionospheric and atmospheric turbulence. We have followed a two-stage self-calibration procedure to correct for residual phase and amplitude errors. This procedure is described in \citet{middelberg2013}. In general terms, the first stage consists of amplitude and phase self-calibration of the data using a model of the phase calibrator and the second step consists of using detected targets with signal-to-noise ratio (S/N) larger than 7 to apply multi-source self-calibration to phase self-calibrate the data. The idea of the multi-source self-calibration is that while individual targets are not sufficiently strong to be used in self-calibration, a combination of the strongest few targets in each epoch in general is. The structure and position of these targets can be divided out using the CLEAN models obtained in imaging. The improvements achieved with this procedure were notable and one example is illustrated in Fig.~\ref{fig:multifield}, where we can see that the peak flux density has increased and the sidelobe level is reduced. After this procedure, we found between five and ten more detected sources per epoch, in comparison to the number of detected sources found before it (a median of $\sim$15 per epoch). \citet{radcliffe2016} tested a calibration algorithm based on multi-source self-calibration on a 1.6\,GHz wide-field VLBI data set of the Hubble Deep Field North and the Hubble Flanking Fields and found great improvements in dynamic range for all the detected sources.
  
\subsubsection{Primary beam correction}

This is a non-standard calibration step. We copied the amplitude and phase corrections from the
  calibration file to the rest of the data sets considering that the
  phase response of a VLBA antenna is constant across the primary
  beam. However, the apparent flux density of a source can be
  attenuated by up to 50\% due to the amplitude response through the
  primary beam. Therefore, we corrected all data sets for primary beam
  attenuation. This correction is described in detail by
  \citet{middelberg2013}, who carried out observations using a pattern
  of pointing positions around 3C\,84 with the VLBA at 1.4\,GHz to
  measure the primary beam response of the antennas. Moreover, we also
  corrected the offset between the beam patterns of the two
  polarizations (VLBA beam squint). We carried out this step using the
  AIPS task CLVLB. 

\subsubsection{Data combination}

This is a non-standard calibration step. Targets in the
  overlap region of pointings have been observed several times. Once
  we considered the calibration of all individual data sets complete,
  the last step was to combine the data of each target observed in
  separate epochs to reach maximum sensitivity. We used the task DBCON
  to combine the calibrated data. External conditions such as weather,
  atmosphere or ionosphere limited the accuracy of the target
  positions. As a result, we observed slight variations in the
  position of the target between epochs. To handle this, we computed
  the median of the variation for each epoch and pointing. To compute the median offset for each epoch we compared the two epochs of the same pointing, taking one of them as reference. To compute the median offset for each pointing we compared each pointing to the reference pointing, which we chose to be the one in the middle of the design (pointing L,
  Fig.~\ref{fig:design}). In every epoch, there were about 450
  targets, of which a median of 20 presented a peak flux
  density exceeding seven times the rms noise. We
  recorded the position of the peak flux density (obtained by fitting a quadratic function to a 3$\times$3 map array) of these 20 targets. To compute the median offset for each pointing, we analysed first the pointings overlapping with
  the reference pointing (pointing L). For each of these pointings, we
  computed the difference of the position of each target which was
  also observed in the pointing L. Then, we examined the pointings
  not overlapping the reference pointing L. We calculated the
  variations considering the ones of the pointing closer to the
  reference pointing, and subsequently added to it. To illustrate this, we will give an
  example for the pointings L-M-N, where we calculated the median
  offset in RA and Dec of pointing M relative to L, and then we added
  the median offset of pointing N relative to M. We made eight detections common to the pointings L and M and recorded the positions of
  each of these detections. The median offset for the pointing M in right ascension ($\Delta$RA$_{M}$) and in declination ($\Delta$Dec$_{M}$) was 0.6 mas and 2.7 mas, respectively. Since there were no
  overlapping regions between the pointing L and the pointing N, we
  used the positions of the pointing M to calculate the variations of
  the pointing N. In this case, we observed ten detections in both pointings M and N. We added the median obtained from the variation in
  right ascension and declination to the variations obtained from
  pointings L-M, resulting in $\Delta$RA$_{N}$=0.6 mas and $\Delta$Dec$_{N}$=1.9 mas. In summary, we used the following equations for this example:
\begin{equation}
\Delta RA_{L} = 0; \Delta Dec_{L} = 0, 
\end{equation}
\begin{equation}
\Delta RA_{M} = RA_{L}-RA_{M}; \Delta Dec_{M} = Dec_{L}-Dec_{M}, 
\end{equation}
\begin{equation}
\Delta RA_{N} = \Delta RA_{M}+(RA_{M}-RA_{N}); \Delta Dec_{N} = \Delta Dec_{M}+(Dec_{M}-Dec_{N}),
\end{equation}
where RA$_{(pointing\_ID)}$ and Dec$_{(pointing\_ID)}$ are the measured positions of the targets in the pointing. Once we
  calculated the variations, we corrected the position of all the
  targets using the task UVFIX. Table~\ref{table:datcomb} contains the median offsets in right ascension and declination for each epoch and pointing. The relative astrometric accuracy of the VLBA is of the order of 10\,micro-arcseconds\footnote{\url{https://science.lbo.us/facilities/vlba}}. Nevertheless, we carried out the phase calibration with the nodding calibration to the phase reference source located $\sim$3\,degrees away. Therefore, the absolute position uncertainties of the sources after phase calibration will be dominated by the residual astrometric error over a few degrees. For any given pointing, the seed positions for multi-source self-calibration is likely to be off by up to a few mas, which is exactly what we see. \citet{deller2016} measured the absolute position uncertainty of the reference position for their sources after phase referencing over a couple of degrees and found it to be $\sim$2\,mas at 1.4\,GHz at a declination of -8\degree. The median of the variations for all the pointings discussed in this project is 0.7 mas in right ascension and 1\,mas in declination. After correcting the positions
  of all the targets, we further measured the peak flux density,
  resulting in a median of 0.02\% increase of the peak flux density
  and three more detections (considering as detection those sources with
  S/N larger than 5.5; for further details see
  Section~\ref{subsec:dets}).


\begin{table}[b!]
\caption{Corrections of the target positions for each epoch and pointing. The corrections of each epoch refer to the offset between the two epochs of the same pointing. The corrections of each pointing refer to the offset between the pointing and the reference pointing L.} \label{table:datcomb}     
\centering                         
\begin{tabular}{| c | r r | r r |}
\cline{2-5}
\multicolumn{1}{c}{} & \multicolumn{2}{|c|}{Offset epoch} & \multicolumn{2}{c|}{Offset pointing}  \\ 
\hline  
Pointing & RA  &  Dec   &  RA  &   Dec \\  
 & (mas) & (mas) & (mas) & (mas) \\
\hline  
A & -0.49 & 2.06 & -0.22 & -0.33 \\
B & -0.04 & 0.68 & 0.83 & 1.29 \\
C & -0.65 & 0.85 & 1.78 & 2.62 \\
D & -1.69 & -1.03 & 1.76 & 0.50 \\
E & -1.42 & 2.60 & 1.40 & 0.52 \\
F & -0.05 & -1.49 & -0.40 & 2.85 \\
G & 0.43 & -3.09 & -0.04 & 1.65 \\
H & 0.51 & 1.60 & -0.30 & 0.26 \\
I & -1.42 & -2.23 & 0.76 & 1.91 \\
J & -1.14 & 2.66 & -0.95 & -0.16 \\
K & -5.97 & -1.16 & 0.51 & 1.01 \\
L & 2.08 & 1.19 & 0.00 & 0.00 \\
M & -0.06 & -1.94 & 0.60 & 2.69 \\
N & -1.94 & -1.10 & 0.61 & 1.91 \\
O & -2.04 & -1.90 & -0.76 & 1.35 \\
P & 0.99 & 2.16 & -0.88 & -1.11 \\
Q & 1.98 & -1.48 & 0.71 & 0.52 \\
R & 0.12 & -0.93 & 0.53 & 1.04 \\
S & -1.65 & -2.92 & 0.55 & 2.14 \\
T & -2.14 & -2.20 & -1.23 & -0.61 \\
U & 3.35 & -2.33 & 1.03 & 0.01 \\
V & 0.25 & -2.18 & 0.82 & 0.17 \\
W & -2.59 & 4.62 & 0.39 & 2.40 \\
\hline                               
\end{tabular}
\end{table}

\begin{figure*}
\centering
\begin{subfigure}{.5\textwidth}
    \centering
    \includegraphics[width=9.5cm]{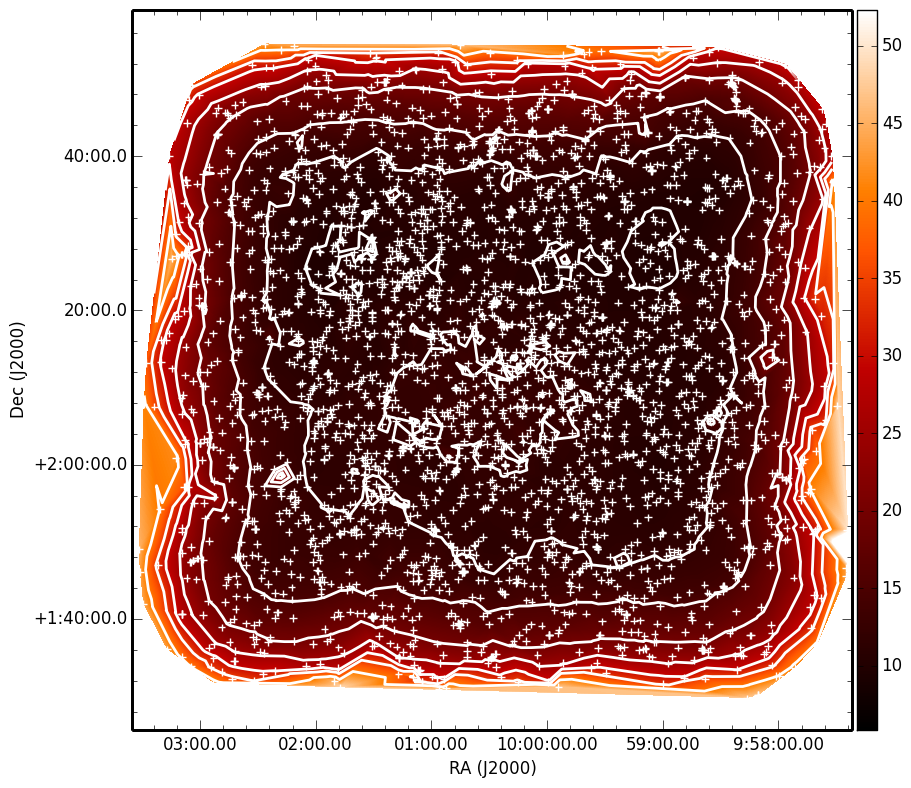}
\end{subfigure}%
\begin{subfigure}{.5\textwidth}
    \centering
    \includegraphics[width=8cm]{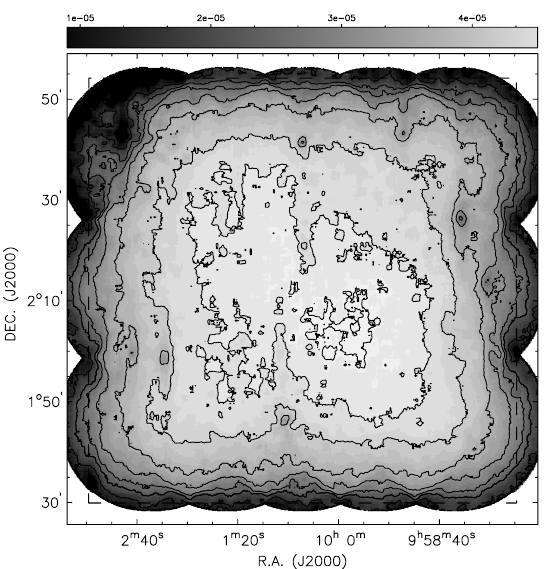}
\end{subfigure}
\caption{\em {Left panel: Sensitivity map of the VLBA-COSMOS
    project. The colour bar represents the rms noise values in
    $\mu$Jy/beam. The white crosses show the target positions. White
    contours are drawn at 10, 12, 15, 20, 25, 30, 34, and 40
    $\mu$Jy/beam to match the contours in the VLA image. Right panel:
    Figure 4 from \citet{schinnerer2010}, representing the sensitivity
    map of the VLA-COSMOS Deep project. The contours correspond to rms
    levels of 10, 12, 15, 20, 25, 30, 34, and 40 $\mu$Jy/beam. We can
    see that the rms distribution of the two images is in excellent
    agreement.}}
\label{fig:sensitmap}
\end{figure*}

\subsubsection{Imaging}

For source detection, we made naturally-weighted images to
  maximise sensitivity using the task IMAGR. Since the VLA
  observations had arcsecond resolution and our VLBA observations have
  milli-arcsecond resolution, we expected the position of the targets
  to be offset from the VLA position. Therefore, we produced big
  images of 4096 x 4096 pixels with a pixel size of 1\,mas. We cleaned
  the data until the first negative Clean component was reached 
    using a clean box rejecting a band of 100 pixels around the image
    edges due to commonly present spurious high values. The median of
  the restoring beam was 16.2 $\times$ 7.3 mas$^{2}$. 
  

For source flux density and position measurements we generated
  uniformly-weighted images. The uniform weighting gives better angular resolution at the expense of sensitivity. Moreover, the distribution of the VLBA antennas produces a plateau in the synthesised beam when natural
  weighting is used, and this plateau was previously found to
  significantly increase the recovered flux density, in particular at
  low S/N (e.g., \citealt{middelberg2013}). We used the task IMAGR as
  explained in the previous step, changing the weighting option to
  uniform. The median of the restoring beam in this case was 12.4
    $\times$ 5.3 mas$^{2}$.

\subsection{Sensitivity map}

For an overview of the final sensitivity of our observations we
computed the rms of all naturally-weighted images. These values were
gridded into an image using linear interpolation between measurements
to cover the region of the COSMOS VLA observations (2
deg$^{2}$). Figure~\ref{fig:sensitmap} shows the sensitivity map
obtained in this project, and the one obtained by
\citet{schinnerer2010} for comparison, since we designed our VLBA observations in order to achieve a similar rms distribution to the VLA observations.

\subsection{Source extraction}

Since the maximum sensitivity is reached in the naturally weighted
images, we used these for source detection. We establish two main conditions to positivily identify detected sources: i) the S/N of the naturally-weighted image should be higher than 5.5 (see Sect.~\ref{subsec:dets}); ii) if the S/N is lower than 7, the position of the VLBA detected source should be within 0\farcs4 of the central part of the VLA radio contours or the optical counterpart (see Sect.~\ref{subsec:counter}). The steps followed to consider a detection as real are explained in detail in Sect.~\ref{subsec:dt}.

\subsubsection{S/N threshold}
\label{subsec:dets}

To minimise the number of false detections, we ran a test using only noise images to establish the S/N threshold for the source detection. We make the noise images by imaging the sky 10\arcsec\ north of the target positions. Taking into account the low probability of finding a source 10\arcsec\, in Declination away from our target, we can estimate the false-detection rate corresponding to several S/N thresholds by measuring the peak flux density of these images. We found that the probability of finding a false-positive together with the probability of having a chance detection within 0\farcs4 of the VLA position is 19\%, 0.2\% and 0.02\% for a S/N threshold of 5, 5.5 and 6, respectively. The false-positive rate for a S/N of 5 is too high (19\%), whereas for a S/N threshold of 6 is almost null (0.02\%) but with the drawback of missing some real detections. Therefore, we decided to consider 5.5 as the S/N threshold, having a false-positive rate of only 0.2\%. 

Considering a S/N of 7, we found a false detection rate of 0.03\% in a 4096x4096 pixel image.



\subsubsection{Radio contours and optical counterparts}
\label{subsec:counter}

\begin{figure*}[t!]
\begin{subfigure}{\textwidth}
    \includegraphics[width=\textwidth, height=8cm]{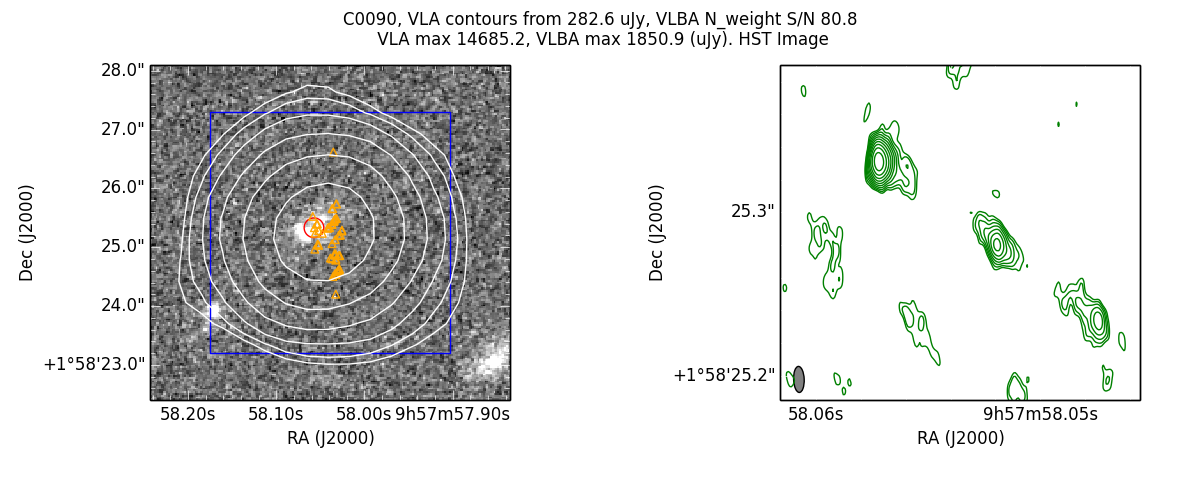}
\end{subfigure}
\begin{subfigure}{\textwidth}
    \includegraphics[width=\textwidth, height=8cm]{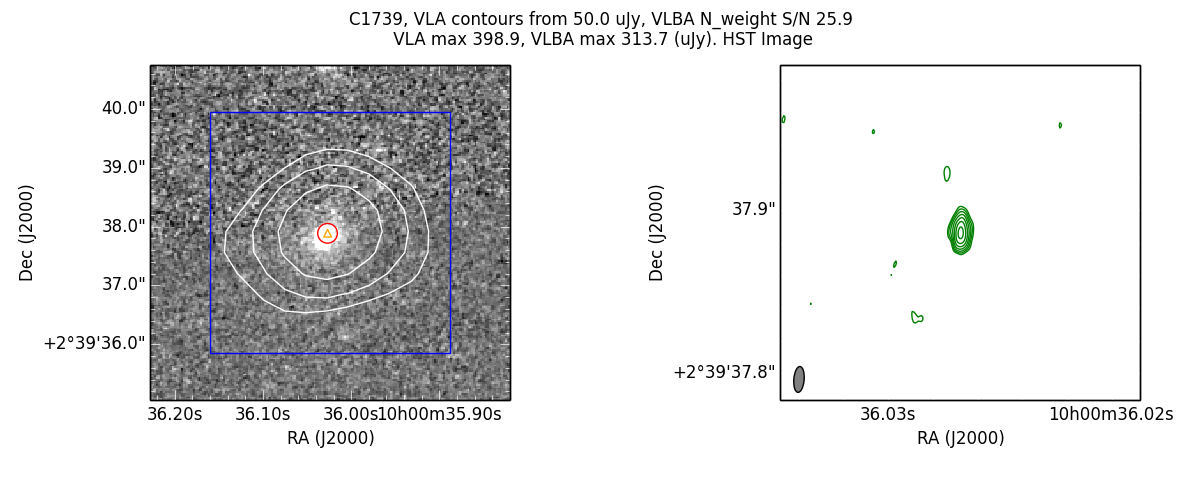}
\end{subfigure}
\caption{\em {Optical counterparts and contour plots of two VLBA
    detections. The header of each pair of panels contains: i) The
    source name used in the present project; ii) the rms noise value
    at which the VLA contours start; iii) The VLBA naturally-weighted
    image S/N; iv) The VLA peak flux density (in $\mu$Jy); v) The VLBA peak flux
    density (in $\mu$Jy); vi) The background greyscale image used (HST or
    Subaru). Left panel: The background greyscale image is the
    HST/Subaru image of the VLBA detection counterpart. The blue
    square represents the 4\arcsec x 4\arcsec VLBA image
    dimension. The white contours represent the VLA contours of the
    source, starting at four times the rms noise level of the VLA
    image and increasing by a factor of two. The red circle represents
    the VLBA peak flux density position. The orange triangles
    represent positions where the S/N of the VLBA naturally-weighted
    image is greater than 5.5. Right panel: Green contours represent
    the VLBA detection contours, starting at three times the rms noise
    level of the naturally-weighted image and increasing by a factor
    of $\sqrt{2}$.}}
\label{fig:count}
\end{figure*}

We created cutout images of 15\arcsec from Hubble Space Telescope Advanced Camera for Surveys (HST-ACS), Subaru (r+ band) and VLA from the NASA/IPAC Infrared Science Archive\footnote{\url{http://irsa.ipac.caltech.edu}} centred on each target position. We plot VLA contours of each target, starting at four times the rms noise level of the VLA image and increasing by a factor of two. Furthermore, we plot VLBA contours, starting at three times the rms noise level of the naturally-weighted image and increasing by a factor of $\sqrt{2}$. Figure~\ref{fig:count} shows some examples of the optical counterparts and the radio contour plots of the targets.

Each pair of panels in Fig.~\ref{fig:count} contains information
about: i) Optical counterpart (when the HST image was not available,
the Subaru image (r-band) was used); ii) VLA contours; iii) Position of the VLBA
peak flux density; iv) VLBA contours of the naturally-weighted image.

\subsubsection{Decision tree}
\label{subsec:dt}

\begin{figure}
\centering
\includegraphics[scale=0.65]{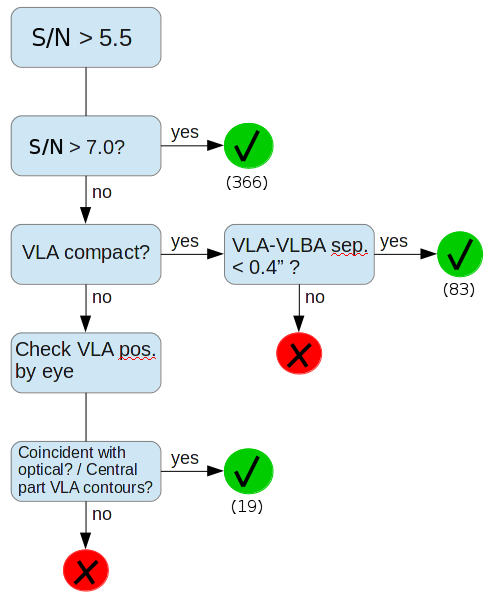}
\caption{\em {Decision tree, through which the detections are passed to check if they are real. The numbers under the green circles correspond to the number of sources fulfilling the related criteria. See text for a detailed explanation of each step.}}
\label{fig:tree}
\end{figure}

To decide if the detections were real, we passed each detection through a decision tree (See Fig.~\ref{fig:tree}). The decision tree was created to minimise human interaction with the data and the probability of having false detections. The steps of the decision tree are as follows:

\def\labelitemi{--}

\begin{figure}
\centering
\includegraphics[scale=0.45]{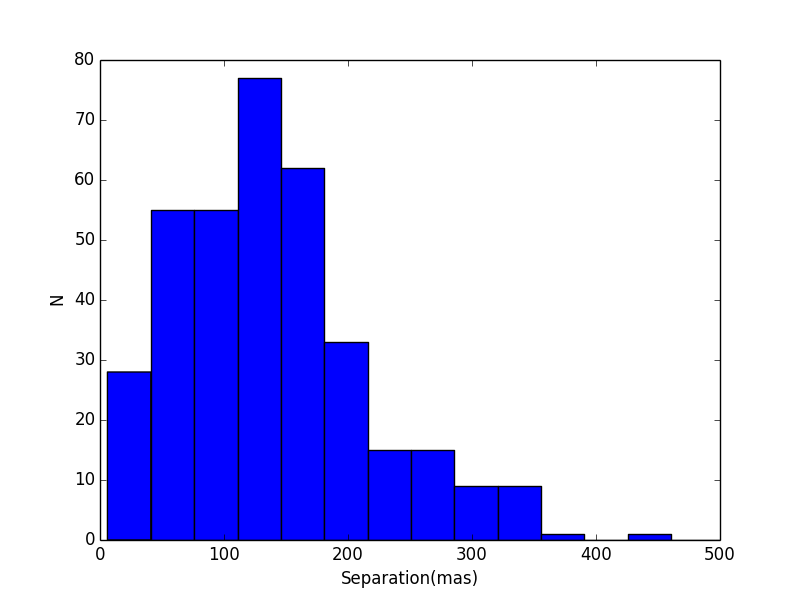}
\caption{\em {Number of VLBA detections with S/N larger than 7 versus
    the separation between the VLA source position and the VLBA source
    position (in milli-arcseconds).}}
\label{fig:histog}
\end{figure}

\begin{enumerate}

\item The first step for a detected source to be passed through the decision tree was that it must present a peak flux density exceeding 5.5 times the local
rms (see Sect.~\ref{subsec:dets}). A total of 710 sources satisfied this criterion.

\item If the S/N was greater than 7, we considered the detection as
  real, since this S/N is high enough to neglect the false detection rate (0.03\%). A total of 366 sources satisfied this criterion.

\item If the S/N was lower than 7, we analysed the compactness of the
  detection. We considered a source compact when it was classified as
  unresolved and single-component source in the VLA catalogue of
  \citet{schinnerer2010}. By plotting the number of detections with
  S/N > 7 versus the VLA-VLBA position separation, we considered that
  a compact source must be located at a separation < 0\farcs4 (See
  Fig.~\ref{fig:histog}), to consider it a real detection. A total of 83 sources satisfied
  this criterion.

\item If the source was considered not compact, then we checked it by
  eye (see Fig.~\ref{fig:count}). If the detection was
  coincident with the optical counterpart we
  considered it a real detection. If no optical counterpart was present,
  and the detection was located in the central part of the VLA
  contours, we considered the detection real. A total of 19 sources
  satisfied this criterion.
\end{enumerate}

After passing our 3293 initial targets through the decision tree, we
ended up with 468 detections. Considering the false detection rate when S/N$>$7 (0.03\%) and when 5.5$<$S/N$<$7 with the VLA-VLBA position separation being smaller than 0\farcs4 (0.2\%), we estimate an overall number of false positives in the final catalogue of $<$1.

\subsection{VLBA flux density and position}
\label{subsec:flxpos}

We ran \texttt{BLOBCAT}\footnote{\url{http://blobcat.sourceforge.net}} on the uniformly-weighted images to measure the flux density and position of the VLBA
detected sources, since, as mentioned in Section~\ref{sec:calib}, the angular resolution of the uniformly-weighted images is better and the plateau of the naturally-weighted images can overestimate the flux density of the source. \texttt{BLOBCAT} catalogues the flood filled islands of pixels (blobs) above a S/N cutoff within a sea of noise, considering each island as a component of a single or multi-component source \citep{hales2012, hales2014}.

We define four parameters before running \texttt{BLOBCAT}: i) \texttt{-\,-pasbe=0.1}, since we consider
the surface brightness (SB) error resulting from calibration to be
10\%; ii) \texttt{-\,-ppe=0.01}, we assume a SB pixellation error of
1\% due to the well-sampled radio images; iii) \texttt{-\,-cpeRA=6.8e-4}; iv) \texttt{-\,-cpeDec=1.98e-3}. These two last parameters define the phase calibrator RA and Dec position errors\footnote{\url{https://www.lbo.us/vlba/astro/calib/}} (in arcsec), respectively. Then, we run \texttt{BLOBCAT} to compute the
position (RA, Dec), its uncertainty ($\Delta$RA, $\Delta$Dec), the rms, the
peak flux density (S$_{p,VLBA}$), its uncertainty ($\Delta$S$_{p,VLBA}$),
the integrated flux density (S$_{i,VLBA}$) and its uncertainty
($\Delta$S$_{i,VLBA}$) for each VLBA detected source. Each output was added to the
catalogue derived from this work (Table~\ref{table:agncat}), after
being checked visually in order to remove the artefacts. We found a median of $\Delta$RA and $\Delta$Dec of 0.8 mas and 2.2 mas, respectively.
 
For multi-component sources a lower case letter was added to the ID of
the source for each component. In these cases, a new line was added
containing the original ID, the weighted average of the position, the rms of the uniformly-weighted image calculated with the
\texttt{AIPS} verb \texttt{imstat}, the sum of the integrated flux
density and its error (Table~\ref{table:agncat}).

The weighted average of the position, p, was calculated as
\begin{equation}
p = \sum (x_{n}\cdot f_{n})/\sum f_{n},
\end{equation}
where {\it {x}} is the position (RA or Dec), {\it {f}} is the flux density and {\it {n}}=1,2...i ({\it {i}} = number of components). 

The error in the sum of the flux density, $\delta$f, corresponds to

\begin{equation}
\delta f = \sqrt{\sum \delta f_{n}^{2}}.
\end{equation}

The high resolution of the VLBA data makes the position of the detected source more precise. Nevertheless, this position still has to match with the VLA position to some degree. Therefore, we compared the positions of the peak flux densities of the VLBA detected sources to the positions of their corresponding VLA targets and we calculated the separation between them. We found a median of the angular separation between the VLA target position and the VLBA detected source of 136 mas.

\begin{figure}
\centering
\includegraphics[scale=0.45]{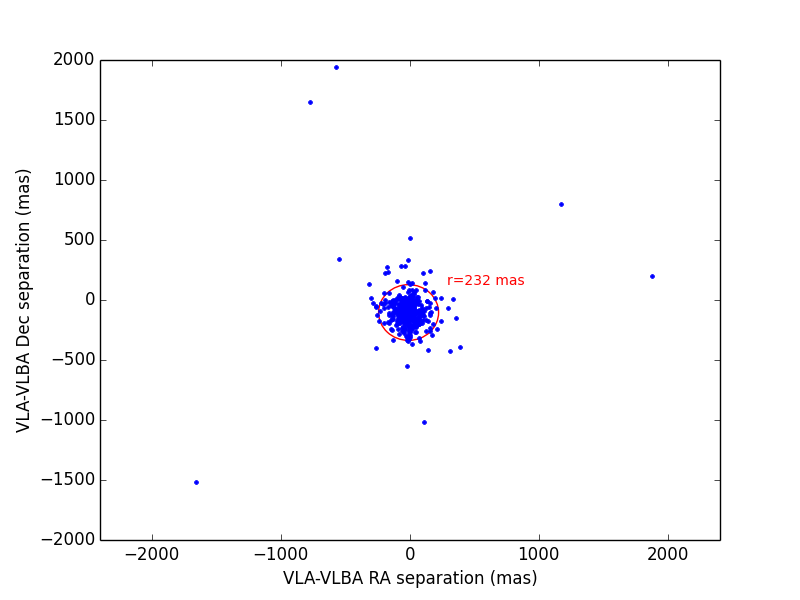}
\caption{\em {Relative positions between VLA
      and VLBA emission. Blue dots show the
    separation in right ascension and declination between the VLA and
    the VLBA source position, in mas. The red circle encompasses the 90\%
    of the detections, with a radius of 232 mas.}}
\label{fig:acc}
\end{figure}

Fig.~\ref{fig:acc} shows the relative positions between the VLA and the VLBA emission. Out of the 468 detections, 421 (90\%) are within a radius of 232 mas offset from the central position of the VLA target. Those detections with an offset greater than 1\arcsec, have a S/N larger than 7, and so are clearly detected, but the VLA source was very extended.

\section{Catalogue}
\label{sec:cat}

We constructed a catalogue containing 468 sources expected to be AGN, of which 14 are considered as multi-component sources (see Table~\ref{table:agncat}). The column entries are the following:

{\it {Column (1)}} -- Source name used in the present project (ID).

{\it {Column (2)}} -- Source name from \citet{schinnerer2010}.

{\it {Column (3)}} -- Integrated VLA flux density of the source (1.4 GHz), in $\mu$Jy, taken from \citet{schinnerer2010}.

{\it {Column (4)}} -- VLBA classification between single- and multi-component source; 0: single-component source; 1: multi-component source.

{\it {Columns (5) and (6)}} -- Right ascension and declination (J2000) of the source, measured with the VLBA (uniform weighting) in degrees.

{\it {Column (7)}} -- Local noise rms measured in $\mu$Jy beam$^{-1}$ with the VLBA (uniform weighting).

{\it {Columns (8) and (9)}} -- Peak flux density of the source and its error, measured in $\mu$Jy beam$^{-1}$ with the VLBA (uniform weighting).

{\it {Columns (10) and (11)}} -- Integrated flux density of the source and its error (see Section~\ref{sec:res} for details), measured in $\mu$Jy with the VLBA (uniform weighting).

We collected complementary multiwavelenth information (see Table~\ref{table:mw}). We considered counterparts within a radius of 1\arcsec. The column entries are the following:

{\it {Column (1)}} -- Source name used in the present project (ID).

{\it {Column (2)}} -- Photometric redshift from \citet{capak2007}, \citet{baldi2014}, \citet{salvato2011}, \citet{kartaltepe2010}, \citet{brusa2010} and \citet{lilly2007}.

{\it {Column (3)}} -- Spectroscopic redshift from \citet{gabor2009}, \citet{trump2009}, \citet{civano2012}, \citet{brusa2010}, \citet{kartaltepe2010b}, \citet{lusso2011}, \citet{lackner2014}, \citet{ranalli2012} and \citet{hao2014}.

{\it {Column (4)}} -- Spitzer/IRAC 3.6 $\mu$m flux density from \citet{brusa2010} and \citet{civano2012}, in $\mu$Jy.

{\it {Column (5)}} -- Spitzer/MIPS 24 $\mu$m flux density from \citet{brusa2010}, \citet{kartaltepe2010} and the Spitzer Enhanced Imaging Products (SEIP) source list from the NASA/IPAC Infrared Science Archive\footnote{\url{http://irsa.ipac.caltech.edu/Missions/spitzer.html}}, in $\mu$Jy.

{\it {Columns (6) and (7)}} -- Soft (0.5-2 keV) and hard (2-10 keV) band fluxes from \citet{civano2016}, \citet{brusa2010}, \citet{cappelluti2009} and \citet{hasinger2007}, in 10$^{-7}$\,W/cm$^{2}$.

{\it {Column (8)}} -- Morphological classification from \citet{tasca2009}, \citet{baldi2014}, \citet{salvato2011}, \citet{trump2009}, \citet{brusa2010}, \citet{gabor2009} and \citet{lusso2011}; 1: early type; 2: spiral; 3: irregular; 4: possible merger; 5: broad emission line object (type 1 AGN); 6: narrow emission line object (type 2 AGN and star-forming galaxies); 7: absorption line galaxies; 8: extended source; 9: compact source; 10: normal/star-forming galaxy; 11: red galaxy; 12: FRI; 13: FRII.

{\it {Column (9)}} -- Stellar mass of the galaxy from \citet{baldi2014}, \citet{kartaltepe2010b} and \citet{lusso2011}, in M$_\odot$.

{\it {Column (10)}} -- Black hole mass from \citet{trump2011} and \citet{hao2014}, in M$_\odot$.

Table~\ref{table:agncat}  and Table~\ref{table:mw} are fragments from the on-line catalogues, available in the electronic edition of the journal\footnote{\label{note1}\url{http://www.aanda.org}}, to illustrate them.

\subsection{False-ID rate}

The cross-matching of our radio catalog with the ancillary data can lead to misidentified sources. Therefore, we have calculated our false identification (false-ID) rate as follows: First, we shifted the positions of all our radio sources by 1\,arcmin both in RA and Dec. Second, we cross-matched our new radio catalog, containing 468 shifted sources, with the catalog from \citet{capak2007}, since it was the one from where we had the highest number of counterparts (389). Finally, we found counterparts for two radio sources, giving a false-ID rate of around 0.4\%.

\begin{table*}
\caption{COSMOS VLBA detections catalogue (1.4 GHz). {\it {Col 1}}: Source name used in the present project; {\it {Col 2}}: Source name from \citet{schinnerer2010}; {\it {Col 3}}: Integrated VLA flux density of the source (1.4 GHz), in $\mu$Jy, taken from \citet{schinnerer2010}; {\it {Col 4}}: VLBA classification between single- and multi-component source, 0: single-component source, 1: multi-component source; {\it {Cols 5, 6}}: Right ascension and declination (J2000) of the source, measured with the VLBA (uniform weighting) in degrees; {\it {Col 7}}: Local noise rms measured in $\mu$Jy beam$^{-1}$ with the VLBA; {\it {Cols 8, 9}}: Peak flux density of the source and its error, measured in $\mu$Jy beam$^{-1}$ with the VLBA (uniform weighting); {\it {Cols 10, 11}}: Integrated flux density of the source and its error (see Section~\ref{sec:res} for details), measured in $\mu$Jy with the VLBA (uniform weighting).}
\label{table:agncat}
\begin{threeparttable}
\centering
\begin{tabular}{lllllllllll} 
\hline\hline             
ID & COSMOSVLADP & S$_{i,VLA}$ & M & RA & Dec &  rms & S$_{p,VLBA}$ & $\Delta$S$_{p,VLBA}$ & S$_{i,VLBA}$ & $\Delta$S$_{i,VLBA}$ \\
  &   &   [$\mu$Jy]  &   & [deg] & [deg] & [$\mu$Jy/ &  [$\mu$Jy/ & [$\mu$Jy/ & [$\mu$Jy] & [$\mu$Jy] \\
  &   &     &   &  &  & beam] &  beam] & beam] &  &  \\  
  (1) & (2) & (3) & (4) & (5) & (6) & (7) & (8) & (9) & (10) & (11)   \\
\hline
  C1641   &   J100028.29+024103.3  &   82740    &       0   &  150.11785    &  2.684271   &  19.7    &       750   &         78        &         833          &  86     \\   
  C1670   &   J100029.62+024018.2  &   326      &       0   &  150.123454   &  2.671707  &   16.5        &   48     &        17        &         73           &  18        \\
  C1679   &   J100030.13+013918.2   &  417        &     0   &  150.125598   &  1.655051  &   18.1     &      116     &       22        &         223          &  29        \\ 
  C1689   &   J100031.10+014044.1   &  379     &        0   &  150.129627   &  1.678869   &  18.4   &        173       &     25       &          214          &  28    \\     
  C1702   &   J100032.43+022845.7   &  279     &        0   &  150.135139   &  2.47937   &   13.3   &        94   &          16        &         145          &  20     \\   
  C1718   &   J100033.99+022645.8   &  142     &        0   &  150.14165    &  2.44605  &    12.4     &      104    &        16        &         119          &  17       \\  
  C1719   &   J100033.99+023905.1  &   290   &          0   &  150.141623   &  2.651423   &  14.2          & 179  &          23        &         183          &  23       \\
  C1722   &   J100034.37+022121.6   &  517     &        0   &  150.143226   &  2.356021    & 12.3   &        67  &           14        &         73           &  14        \\
  C1725   &   J100034.83+014247.2   &  345        &     0   &  150.145154   &  1.713091   &  17.5   &        186       &     26        &         205          &  27       \\
  C1739   &   J100036.02+023937.9  &   539     &        0   &  150.150112   &  2.660524   &  15.3    &       290      &      33       &          337          &  37         \\
  C1740   &   J100036.05+022830.6  &   530    &         0  &   150.150222   &  2.475156   &  13.9        &   85    &         16       &          100          &  17       \\ 
  C1750   &   J100037.65+022949.0  &   187    &         0  &   150.156882   &  2.496944   &  12.7   &        152    &        20         &        133          &  18     \\    
  C1758   &   J100038.25+022327.8   &  114     &        0   &  150.159364   &  2.391057  &   12.4      &     56          &   14        &         111          &  17      \\  
  C1763   &   J100038.45+024157.6   &  278     &        0   &  150.160232   &  2.699306   &  18.4     &      106   &         21       &          127          &  22    \\   
  C1769   &   J100039.27+015243.7  &   115     &        0   &  150.163665   &  1.878767  &   13.9     &      58         &    15       &          61           &  15      \\  
  C1773   &   J100039.96+023118.2   &  132        &     0   &  150.166518   &  2.521681   &  12.6    &       71  &           14       &          72           &  15         \\
  C1774  &    J100040.00+023131.0  &   55    &          0    & 150.166629   &  2.525287   &  12.4         &  61   &          14      &          55           &  14      \\ 
  C1779   &   J100040.86+020431.2  &   131   &          0    & 150.170255   &  2.075272  &   14.4    &       97   &          17       &          77           &  16      \\ 
  C1780   &   J100040.91+021307.7  &   213        &     0   &  150.170455   &  2.218779   &  14.1      &     66    &         16   &              75           &  16      \\  
  C1781  &    J100041.16+020502.7  &   203      &       0   &  150.171474   &  2.084064  &   14.0     &      65    &         15           &      92           &  17      \\  
  C1784  &    J100041.41+023124.1  &   716     &        0  &   150.172571   &  2.523339   &  14.1      &     93    &         17     &            127          &  19      \\   
  C1798  &    J100042.39+020939.8   &  251  &           0   &  150.176651   &  2.161034   &  14.6   &        109  &          18              &   98           &  18       \\ 
  C1810  &    J100043.17+014607.9   &  88170     &      0   &  150.179973   &  1.768862  &   19.3    &       1544         &  156      &          2525         &  253      \\  
  C1819   &   J100043.53+022524.4   &  589         &    0  &   150.181401   &  2.423415    & 12.3  &         173  &          21       &         206          &  24      \\  
  C1824   &   J100044.55+013942.2   &  1803     &       0    & 150.185629   &  1.661681    & 20.1    &       898  &          92       &          1150         &  117        \\
 C1833   &   J100045.25+015459.0  &   139       &      0   &  150.188504   &  1.916319   &  14.0    &       71   &          16      &           86           &  16       \\  
 C1847   &   J100045.80+020119.0   &  476    &         0  &   150.190843   &  2.021931   &  15.7     &      278  &          32   &              407          &  44        \\ 
 C1860   &   J100046.91+020726.5   &  2204        &    1   &  150.195473   &  2.124031   &  14.8    &           &             &            1383         &  96         \\
 C1860a  &   J100046.91+020726.5   &  2204   &         0   &  150.195469   &  2.124033  &   19.4     &      581      &      62     &            783          &  81        \\
 C1860b  &   J100046.91+020726.5   &  2204    &        0  &   150.195476   &  2.12403   &   20.5     &      193       &     28        &         373          &  43         \\
 C1860c  &   J100046.91+020726.5   &  2204          &  0  &   150.195485   &  2.124025   &  18.9  &         146        &    24  &               227          &  30        \\ 
 C1875   &   J100047.60+015910.3   &  21470    &       0  &   150.198312   &  1.986288   &  17.7    &       233  &          29           &      290          &  34        \\
   C1884    &   J100048.53+013914.0   &   147       &   0   &   150.202244   &   1.653855   &   19.7    &   144     &   24        &   144     &   22    \\
  C1886    &   J100048.89+023127.5   &   234       &   0   &   150.203709   &   2.52428    &   12.3    &   122     &   17        &   140     &   19    \\
  C1893    &   J100049.58+014923.7   &   15100     &   0   &   150.20663    &   1.82326    &   16.5    &   669     &   69        &   835     &   85    \\
  C1895    &   J100049.65+014048.9   &   243       &   0   &   150.20689    &   1.680226   &   18.7    &   76      &   20        &   95      &   21    \\
  C1896    &   J100049.78+021654.9   &   1098      &   0   &   150.20742    &   2.281892   &   14.9    &   512     &   54        &   713     &   73    \\
  C1897    &   J100049.91+020500.0   &   311       &   0   &   150.207967   &   2.083336   &   14.8    &   111     &   19        &   236     &   28    \\
  C1903    &   J100050.45+023356.1   &   610       &   0   &   150.210223   &   2.565572   &   12.7    &   323     &   35        &   396     &   42    \\
  C1911    &   J100051.21+014027.3   &   996       &   0   &   150.21342    &   1.674212   &   18.6    &   149     &   24        &   149     &   22    \\
  C1938    &   J100054.59+020459.5   &   121       &   0   &   150.227436   &   2.08317    &   13.4    &   114     &   18        &   114     &   17    \\
  C1949    &   J100055.36+015955.2   &   124       &   0   &   150.230691   &   1.998687   &   13.2    &   60      &   15        &   110     &   17    \\
  C1959    &   J100056.08+014347.3   &   615       &   0   &   150.233688   &   1.729789   &   17.7    &   86      &   20        &   86      &   19    \\
  C1975    &   J100057.06+022942.9   &   123       &   0   &   150.237771   &   2.495213   &   12.9    &   108     &   17        &   108     &   16    \\
  C1977    &   J100057.11+023451.7   &   347       &   0   &   150.237968   &   2.581038   &   14.1    &   233     &   27        &   240     &   28    \\
  C1978    &   J100057.16+013217.8   &   252       &   0   &   150.238212   &   1.538304   &   42.1    &   197     &   46        &   481     &   64    \\
  C1983    &   J100057.33+020839.0   &   193       &   0   &   150.238855   &   2.144165   &   15.3    &   82      &   17        &   105     &   19    \\
  C1988    &   J100057.45+024217.1   &   1047      &   0   &   150.239409   &   2.704769   &   17.9    &   426     &   46        &   454     &   49    \\
  C1995    &   J100057.94+015819.3   &   319       &   0   &   150.241416   &   1.972019   &   13.5    &   65      &   15        &   101     &   17    \\
  C2002    &   J100058.05+015129.0   &   13260     &   0   &   150.242241   &   1.859517   &   14.1    &   85      &   17        &   164     &   22    \\
\hline
\end{tabular}
\begin{tablenotes}
\small
\item Fragment from the on-line catalogue, available in its entirely in the electronic edition of the journal, to be used as guidance of the content.
\end{tablenotes}
\end{threeparttable}
\end{table*}

\begin{table*}
\caption{VLBA-COSMOS multi-wavelength counterparts. {\it {Column (1)}}: Source name used in the present project; {\it {Column (2)}}: Photometric redshift; {\it {Column (3)}}: Spectroscopic redshift; {\it {Column (4)}}: Spitzer/IRAC 3.6 $\mu$m flux density, in $\mu$Jy; {\it {Column (5)}}: Spitzer/MIPS 24 $\mu$m flux density, in $\mu$Jy; {\it {Columns (6) and (7)}}: Soft (0.5-2 keV) and hard (2-10 keV) band fluxes, in 10$^{-7}$\,W/cm$^{2}$; {\it {Column (8)}}: Morphological classification, 1: early type, 2: spiral, 3: irregular, 4: possible merger, 5: broad emission line object, 6: narrow emission line object, 7: absorption line galaxies, 8: extended source, 9: compact source, 10: normal/star-forming galaxy, 11: red galaxy, 12: FRI, 13: FRII; {\it {Column (9)}}: Stellar mass of the galaxy, in M$_\odot$; {\it {Column (10)}}: Black hole mass, in M$_\odot$. For references see details in Section~\ref{sec:cat}.}             
\label{table:mw}     
\begin{threeparttable}
\centering                         
\begin{tabular}{| l | l | l | l | l | l | l | l | l | l |}      
\hline\hline           
ID & zphot & zspec & F3.6 & F24 & SFlux(0.5-2 keV) & HFlux(2-10 keV) & Mph & logM* & logMBH \\    
  &   &    & [$\mu$Jy]  & [$\mu$Jy] & [10$^{-15}$ 10$^{-7}$\,W/cm$^{2}$] & [10$^{-15}$ 10$^{-7}$\,W/cm$^{2}$] &  & [Msun] &  [Msun]  \\
  (1) & (2) & (3) & (4) & (5) & (6) & (7) & (8) & (9) & (10)  \\
\hline                        
  C1641   &   0.32   &   0.3493  &   323.56   &   131.5   &   4   &   6.1 &   1    &    11.46  &             \\
  C1670   &   0.33   &         &          &             &    0.7     &     3.7              &  1    &          &              \\
  C1679   &   1.35   &         &           &          &            &                  &  2    &           &             \\
  C1689   &   0.94   &   0.84   &          &            &          &                  &  1    &         &               \\
  C1702   &        &             &        &      140.2   &           &                  &        &         &              \\
  C1718   &   0.93   &   0.9    &           &          &          &                   &  1     &      &                 \\
  C1719   &   1.22  &         &           &              &    0.5      &  2.7               &  2     &         &              \\
  C1722   &       &           &           &              &        &                   &        &          &             \\
  C1725   &       &           &            &      638.5   &            &                 &       &         &               \\
  C1739   &   1.38    &        &               &        &         &                   &  2     &         &              \\
  C1740   &   0.79  &    0.6879    & 286.5    &   3876.0  &    3.3    &  50    &  2     &   11.31   &  9.48      \\
  C1750    &  0.84    &  0.671  &    30.25   &         &   0.6      & 9.3                   &  1     &         &              \\
  C1758   &   0.98   &          &           &           &         &                   &  2      &        &              \\
  C1763   &   0.3    &   0.72   &           &     702.7 &         &                    &  2      &  11.32    &           \\
  C1769  &    0.64   &      &             &             &         &                   &  1     &         &              \\
  C1773   &   0.82   &        &            &           &          &                   &  1     &         &              \\
  C1774   &   0.74    &      &            &            &          &                   &  1     &        &               \\
  C1779    &         &          &          &            &             &               &        &        &               \\
  C1780   &   1.05  &    1.156   &   10.8    &    202.2   &   1.6   &   8.1   &   2  &            &              \\
  C1781   &   0.73   &         &             &         &            &                 &  2     &        &               \\
  C1784   &   0.76   &         &            &          &           &                  &  1      &         &             \\
  C1798   &   2.12   &         &          &             &         &                   &         &       &               \\
  C1810    &  0.34  &    0.346    &  224.72  &    100.0     &  3.5  &   7.5 &   1    &    10.99   &            \\
  C1819   &   0.84   &   0.7274   &  46.93     &  120.0    &  0.3  &   5.1 &   1    &    10.75   &  9.29      \\
  C1824   &   0.17   &       &            &             &        &                    &  1     &       &                 \\
  C1833   &   1.03   &           &          &            &         &                  &  2     &           &            \\
  C1847   &   1.05   &         &           &            &           &                 &  3     &       &                \\
  C1860    &  1.73   &   1.158  &           &     7322.0  &          &                  &  2     &   11.31  &             \\
  C1875   &   0.41    &  0.438  &    145.66  &            &    0.7   &  4.3                 &  1    &          &              \\
  C1884   &   0.17    &            &             &           &                    &                   &   1     &           &            \\
  C1886   &   1.28    &            &             &           &                    &                   &   2     &           &            \\
  C1893   &   0.53    &   0.53     &             &           &                    &                   &   1     &           &            \\
  C1895   &   1.32    &   0.7134   &   108.44    &   632     &   1.6         &   13.3        &   3     &           &            \\
  C1896   &   0.94    &   0.88     &             &           &                    &                   &   1     &   10.08   &            \\
  C1897   &   0.45    &   1.2373   &   190.11    &   3052    &   55         &   75        &   1     &           &   8.17     \\
  C1903   &   1.46    &            &             &   393     &                    &                   &   2     &           &            \\
  C1911   &   0.22    &   0.166    &             &   7850    &                    &                   &   1     &   11.1    &            \\
  C1938   &   0.98    &            &             &           &                    &                   &   2     &           &            \\
  C1949   &   1.73    &   2.22     &   10.98     &   812     &   0.4        &   2.5      &   1     &           &            \\
  C1959   &           &            &             &   730     &                    &                   &         &           &            \\
  C1975   &   2.58    &            &             &           &                    &                   &   1     &           &            \\
  C1977   &           &            &             &           &   1.1     &   3.1       &         &           &            \\
  C1978   &   1.36    &            &             &   180     &                    &                   &         &           &            \\
  C1983   &   1.73    &            &             &           &                    &                   &   2     &           &            \\
  C1988   &   1.27    &            &             &           &                    &                   &   2     &           &            \\
\hline                                  
\end{tabular}
\begin{tablenotes}
\small
\item Fragment from the on-line catalogue, available in its entirely in the electronic edition of the journal, to be used as guidance of the content.
\end{tablenotes}
\end{threeparttable}
\end{table*}



\section{Results and discussion}
\label{sec:res}

\begin{figure}
\includegraphics[scale=0.4]{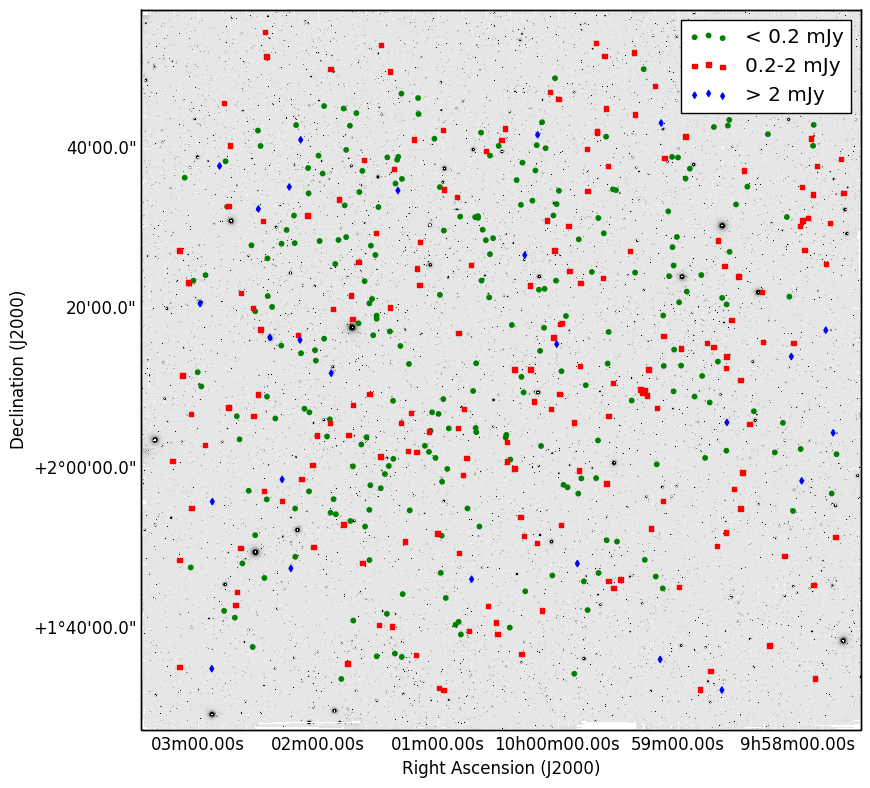}
\caption{\em {Distribution of the 468 VLBA detected sources over the COSMOS field. The green circles, red squares and blue diamonds represent VLBA detections whith S$_{int}$ < 0.2 mJy, 0.2 mJy < S$_{int}$ < 2 mJy and S$_{int}$ > 2 mJy, respectively. The background greyscale image is a mosaic of the COSMOS Subaru i-band data.}}
\label{fig:detcos}
\end{figure}

The positions in the COSMOS field of the 468 sources detected by the VLBA are shown in Fig.~\ref{fig:detcos}. We can see that the distribution of the sources is roughly homogeneous.

\subsection{Detection fraction}

In the case of VLA multi-components sources detected with the VLBA, only the core of the source was detected. This is reasonable, since the rest of the components are expected to be extended regions, i.e., less likely for a VLBA detection. We searched then for the number of the radio sources from the input catalogue, which would be detectable by the VLBA. We considered a VLA source as detectable by the VLBA when its peak flux density exceeds 5.5 times the local noise level of the VLBA naturally weighted image. Applying this criterion, we found that 2361 out of the 2865 VLA sources were in principle detectable. For this reason, we evaluate the detection fraction based on these 2361 sources.

\begin{figure}
\centering
\includegraphics[width=9.8cm, height=8.2cm]{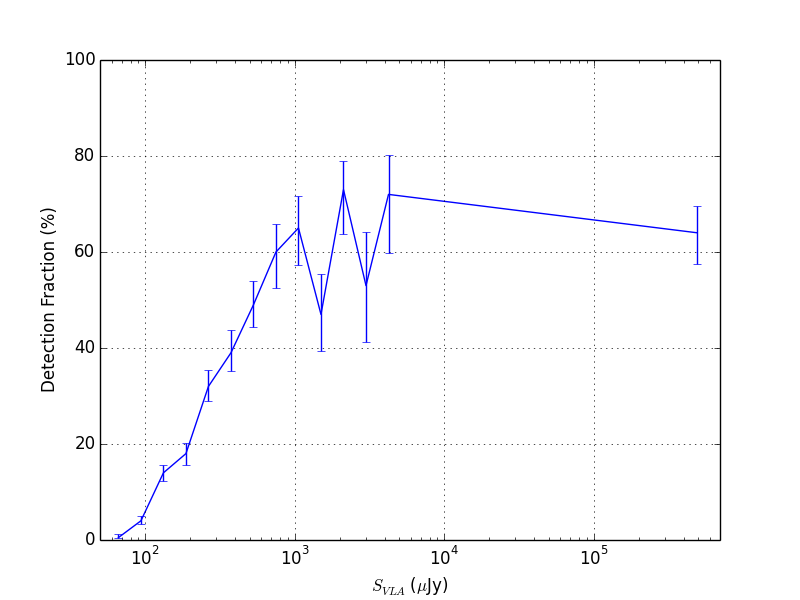}
\caption{\em {VLBA detection fraction as a function of S$_{VLA}$. Only the VLBA detectable sources have been considered to compute this fraction. The VLA flux densities have been separated into bins corresponding to 55$\mu$Jy\,$\times$\,$\sqrt{2^{N}}$, where N=0,1,2,.... The error bars have been calculated using the Bayesian technique for binomial populations as described by \citet{cameron2011} (1\,$\sigma$ confidence interval).}}
\label{fig:det}
\end{figure}

We computed the detection fraction as a function of the VLA flux density (S$_{VLA}$) (see Fig.~\ref{fig:det} and Table~\ref{table:dett}). The uncertanties shown here have been calculated using the Bayesian beta distribution quantile technique as described by \citet{cameron2011} (1\,$\sigma$ confidence interval). It can be seen that, at flux densities $<$1\,mJy, the detection fraction is higher for higher flux densities. At higher flux densities, the detection fraction is independent of flux density. These numbers give us a lower limit on the abundance of AGNs in the field, and in particular at faint flux densities, where the achieved sensitivity and the source compactness play a more important role. At high flux densities, a source could be detected even if the VLBI core accounts for a small percentage of the total flux density.

\begin{figure}
\centering
\includegraphics[scale=0.4]{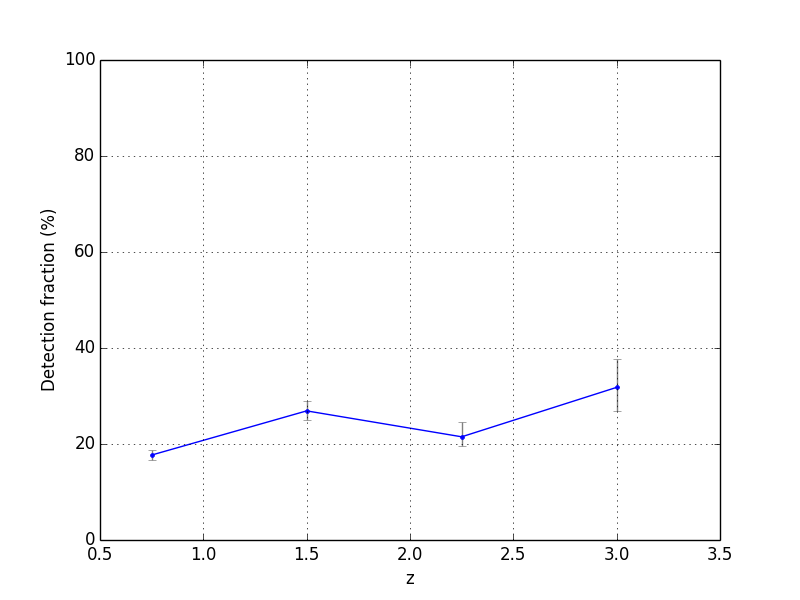}
\caption{\em {VLBA detection fraction as a function of redshift. Photometric redshifts have been used (Sect.~\ref{sec:cat}). The error bars have been calculated using the Bayesian technique for binomial populations as described by \citet{cameron2011} (1\,$\sigma$ confidence interval).}}
\label{fig:detz}
\end{figure}

Fig.~\ref{fig:detz} shows the detection fraction as a function of redshift. Photometric redshifts have been used for the redshift bins (see Sect.~\ref{sec:cat}). It can be seen that the detection fraction is roughly constant over the redshift range $0.5\,<\,z\,<\,3$, showing no evolution. This is in agreement with the findings from \citet{rees2016}. The slight rise of the detection fraction at the highest redshifts probably is due to the small VLBA sample at these redshifts, which make the uncertainties higher.

Table~\ref{table:detf} shows the comparison between the number of VLBI detections from various projects (including the present project), grouped into four flux density bins. The Chandra deep field south (CDFS) project achieved a 1\,$\sigma$ sensitivity of 55\,$\mu$Jy\,beam$^{-1}$ and detected 20$^{+5}_{-4}$\% of the sources \citep{middelberg2011}. The Lockman Hole/XMM project achieved a 1\,$\sigma$ sensitivity of 24\,$\mu$Jy\,beam$^{-1}$ and detected 30$\pm$3\% of the sources \citep{middelberg2013}. The mJIVE project \citep{deller2014}  achieved a 1\,$\sigma$ sensitivity of 60\,$\mu$Jy\,beam$^{-1}$ and detected 20$\pm$0.3\% of the sources. In the present project we  achieve a 1\,$\sigma$ sensitivity of 10\,$\mu$Jy\,beam$^{-1}$ and detect 20$\pm$1\% of the sources (468 detections out of 2361 detectable sources).

\begin{table}
\caption{VLBA detection fraction as a function of the VLA flux density.} \label{table:dett}     
\centering                         
\begin{tabular}{c c c c}        
\hline\hline                 
Flux density bin & N$_{VLA}$ & N$_{det}$ & Det. fract. \\   
($\mu$Jy) &  &  & (\%) \\   
\hline                        
  55.0-77.8  & 382 & 2 & 0.5 \\     
  77.8-110.0  & 578 & 23 & 4 \\
  110.0-155.6  & 435 & 60 & 14 \\
  155.6-220.0  & 265 & 47 & 18 \\
  220.0-311.1  & 203 & 65 & 32 \\ 
  311.1-440.0  & 132 & 52 & 39 \\     
  440.0-622.3 & 106 & 52 & 49 \\
  622.3-880.0  & 52 & 31 & 60 \\
  880.0-1244.5  & 43 & 28 & 65 \\
  1244.5-1760.0  & 36 & 17 & 47 \\ 
  1760.0-2489.0  & 33 & 24 & 73 \\     
  2489.0-3520.0  & 17 & 9 & 53 \\
  3520.0-4978.0  & 18 & 13 & 72 \\
  $>$4978.0  & 61 & 39 & 64 \\
\hline                               
\end{tabular}
\end{table}

The targeted sources by the CDFS, Lockman Hole/XMM and mJIVE-20 projects were mainly bright sources, whereas the targeted sources by the VLBA-COSMOS project were mainly faint sources. Since it is assumed that brighter sources are more likely to hold an AGN, the percentage of detected sources with the former three projects are expected to be higher than with our VLBA-COSMOS project. Nevertheless, the sensitivity of our project is much better than those of the other three projects, making the probability to detect a source higher. Therefore, the detection fractions of the four projects are roughly in good agreement with the expectations. One possible explanation of the slightly higher detection fraction of the Lockman Hole/XMM project is the combination of targeting relatively bright sources with the better achieved sensitivity than those achieved by the CDFS and mJIVE-20 projects.

\begin{table}
\caption{Number of VLBA detected radio sources by this project, grouped into four flux density bins, compared with the number of detections of different projects. References: D14 - \citep{deller2014}; M13 - \citet{middelberg2013}; M11 - \citet{middelberg2011}.}
\label{table:detf}     
\centering                         
\begin{tabular}{c c c c c c}        
\hline\hline                 
 &  & 100 & 0.5 & 1 & \\   
 Project & <100 & -500 & -1 & -10 & ref. \\   
 & $\mu$Jy & $\mu$Jy & mJy & mJy & \\   
\hline                        
   COSMOS & 114 & 246 & 57 & 45 & \\     
   mJIVE & 0 &  8   & 307 & 3679 & D14 \\ 
   Lockman & 0 &  35   & 12 & 17 & M13 \\
   CDFS & 0 &   1   &  3 & 16 & M11 \\
\hline                               
\end{tabular}
\end{table}

\subsection{VLBA-VLA flux density ratio}

\begin{figure}
\includegraphics[width=9.8cm, height=8.2cm]{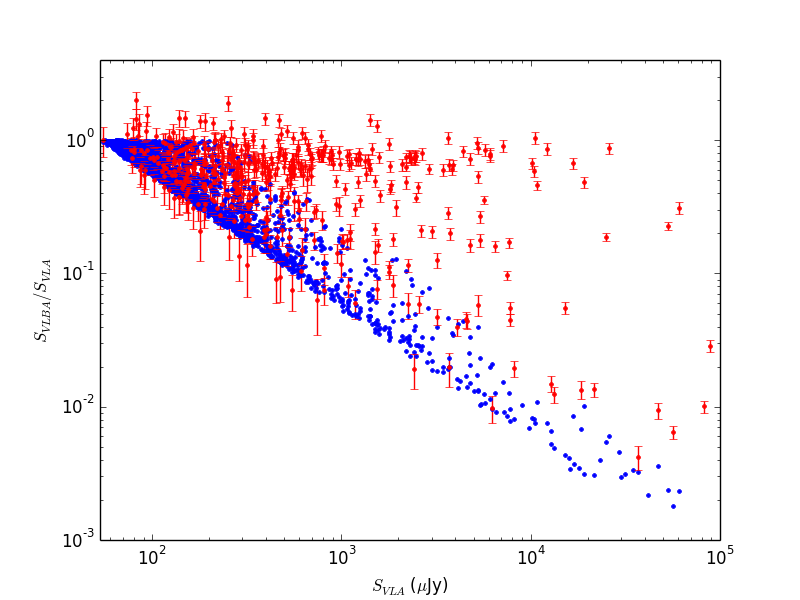}
\caption{\em {VLBA-VLA flux density ratio. The blue dots represent upper limits for the VLA sources from the input catalog. Their VLBA-VLA flux density ratios have been computed as $5.5\times\,rms_{VLBA}/S_{VLA}$, where $rms_{VLBA}$ is the rms noise of the VLBA naturally-weighted image (since we used the natural weighting for source detection), and $S_{VLA}$ is the integrated VLA flux density of the source. The red dots with error bars represent the VLBA detected sources.}}
\label{fig:hstf}
\end{figure}

Fig.~\ref{fig:hstf} exhibits a diagram showing the VLBA-VLA flux density ratio as a function of the integrated VLA flux density. The median value for the VLBA-VLA flux density ratio of the VLBA detected sources is 0.6. At high VLA flux densities ($S_{VLA}>$10\,mJy), the VLBA-VLA flux density ratio appears to be spread roughly uniformly between 0.01 and 1. At low VLA flux densities ($S_{VLA}\lesssim$1\,mJy), it seems that there is an overpopulation of VLBA detected sources with recovered flux densities between 60\% and 80\% of their VLA flux density, suggesting that a large number of low flux density sources detected with the VLBA not only have an AGN but they are also dominated by it. In particular, 237 out of the 344 VLBA detected sources with $S_{VLA}\lesssim$1\,mJy (69\%) have more than half of their total radio luminosity in a VLBI-scale component, whereas this is true for only six of the 21 VLBA detected sources with $S_{VLA}\gtrsim$10\,mJy (29\%).

To homogenise the samples we should eliminate the bias caused by the fact that at faint flux densities only compact sources can be detected. In order to do so, we considered only sources where the sensitivity of the VLBA observations would allow us to detect a VLBA component (i.e., more than half of their total radio luminosity recovered with the VLBA). In this case, the percentage of sources with $S_{VLA}\lesssim 1$\,mJy and more than half of their total radio luminosity in a VLBI-scale component shows only a small variation from 69\% to 64\%.

These results suggest that low flux density sources have a greater fraction of their radio luminosity in the core and that stronger sources have more extended emission. This is in agreement with the findings from \citet{deller2014}, which showed that fainter sources were somewhat more likely to be dominated by a very compact component than brighter sources. \citet{mullin2008} also found a decrease of core prominence with source luminosity and proposed that higher luminosity sources, with faster jets, experience stronger Doppler suppression as an explanation.

Another explanation of why the faint and bright AGN populations differ is that the stronger sources may have large-scale jets and lobes, thus only a smaller fraction of the flux is from the core, whereas the fainter ones tend to be compact. Only 0.9\% of the VLBA detected sources with $S_{VLA}\lesssim 1$\,mJy are classified as multi-component in the VLA catalogue of \citet{schinnerer2010}, while this is the case for 50\% of the VLBA detected sources with $S_{VLA}\gtrsim 10$\,mJy.

The observed difference may also be a consequence of age effects. Many of the faint compact sources presumably are objects similar to the gigahertz peaked-spectrum (GPS) sources, which are thought to be very young and to evolve into the large-scale Fanaroff-Riley class I (FR\,I) and class II (FR\,II) radio galaxies \citep{tinti2006}. However, the reason behind the high number of observed GPS sources compared to the low number of observed FR\,I and FR\,II sources is still not well understood.

It is also worth noting that composite galaxies (with both AGN and star formation playing an important role) are claimed to be a part of the faint radio population (e.g. \citealt{strazzullo2010}). Therefore, it is expected that the ratio of compact sources actually drops for the faint sample since the star-formation related radio emission should be resolved out by VLBI. This is the opposite of what we see since we find more compactness in the faint sample of VLBA detected sources, suggesting that the difference between the faint and the bright AGN populations must be even more pronounced because some of it is being masked by the starburst sources.

4\% of the sources (19/468) have a $S_{VLBA}$ larger than $S_{VLA}$ by more than 1\,$\sigma$ (the length of their error bars). Some of these sources might have larger flux density ratios because of normal Gaussian errors. Three sources out of the 468 (0.6\%) have a $S_{VLBA}$ larger than $S_{VLA}$ by more than 3\,$\sigma$. One possible explanation to find VLBA flux densities larger than the VLA flux densities is variability. We note that the VLA observations were performed between 2004 and 2006, while the VLBA observations were performed between 2012 and 2013. Moreover, these sources are towards fainter sources, whose likelihood to be detected increases if they are in a high flux density state.


\subsection{Detected and undetected sources}

We make use of our results and the complementary information in the literature to study the properties of the detected and undetected sources. 

\subsubsection{Redshifts}

We found spectroscopic redshifts of 129 out of 468 VLBA detected sources (28\%), and 229 out of 2397 VLBA undetected sources (10\%) (see Section~\ref{sec:cat} for the references). The maximum redshift reached by the detected sources is 3.1, and by the undetected sources is 2.8. The median redshift of the detected and undetected sources are 0.71 and 0.59, respectively. The number of available spectroscopic redshifts is still low. Nevertheless, the available photometric redshifts of the COSMOS field have been demonstrated to have a high accuracy (e.g. \citealt{salvato2011}) and therefore are highly reliable.

\begin{figure}
\includegraphics[width=9.8cm, height=8.2cm]{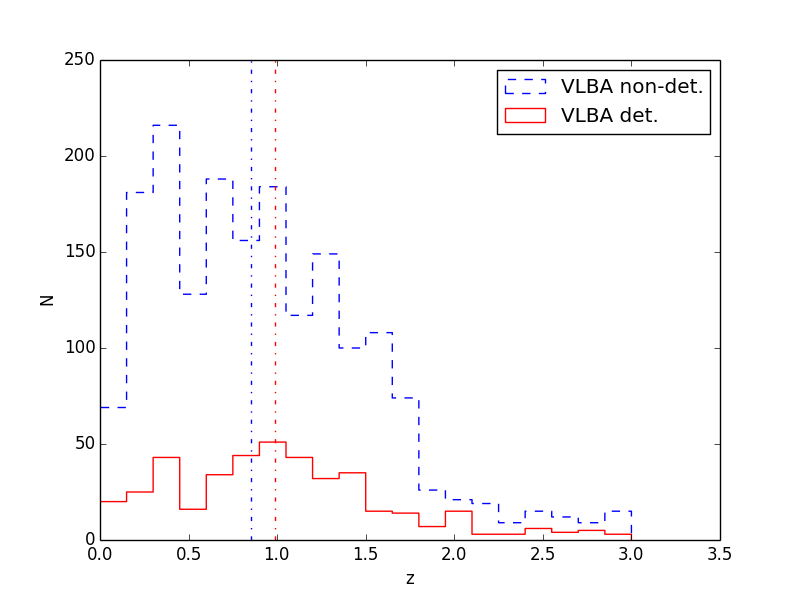}
\caption{\em {Number of VLBA detected (red histogram) and undetected sources (blue dashed histogram) as a function of redshift. The vertical dot-dashed lines represent the median photometric redshifts of each sample.}}
\label{fig:histz}
\end{figure}

We have recovered photometric redshifts for 413 out of 468 VLBA detected sources (89\%), and 1799 out of 2397 VLBA undetected sources (75\%) (see Section~\ref{sec:cat} for the references). The maximum redshift achieved by the detected and undetected sources is, in both cases, 3. The median redshift in the case of detected sources is 0.99, and in the case of undetected sources is 0.85. Figure~\ref{fig:histz} shows the redshift distributions of the VLBA detected and undetected sources.

The median redshift of the VLBA detected sources of $\sim$1 fortifies the premise that they likely hold an AGN since the brightness temperature of a source to be detected with VLBI observations at that redshift can only be reached by the AGN activity.

We note that even if the total flux density of a source is high, it could not be detected with a high-resolution interferometer if the surface brightness of the source is low. Therefore, one possible explanation of the median redshift of VLBA undetected sources being lower than the median redshift of the VLBA detected sources is that they have low brightness temperatures.

\subsubsection{VLA flux densities}

The median VLA peak flux density of the VLBA detected sources is 320 $\mu$Jy, with a maximum of 24.5\,mJy and a minimum of 55 $\mu$Jy, whereas the median VLA peak flux density of the VLBA undetected sources is 100 $\mu$Jy, being the maximum 19.1\,mJy and the minimum 27 $\mu$Jy. The low flux density of the majority of undetected sources is in agreement with our expectations due to i) a changing underlying source population at lower flux density and ii) the higher need of a very compact core for the fainter sources to be detected.

If we analyse the VLA integrated flux density of the VLBA detected and undetected sources we obtain similar values for the median, 390 $\mu$Jy and 120 $\mu$Jy, respectively. In this case, the maximum flux density is 88.17\,mJy for the detected sources and 175.5\,mJy for the undetected sources. The non-detection of the VLA sources with the highest integrated flux densities can be explained as in the previous subsection, since the VLBA could not detect a source of low surface brightness, regardless of its total flux density.


\subsubsection{X-ray fluxes}
\label{sec:xra}

We have found X-ray counterparts in the soft band (0.5-2 keV) for 136 VLBA detected sources (29\%) and 341 undetected sources (14\%), without considering upper limit values. In the case of X-ray fluxes in the hard band (2-10 keV), we found counterparts for 132 VLBA detected sources (28\%) and 337 undetected sources (14\%) (see Section~\ref{sec:cat} for the references). 

The median value for the soft band fluxes of the detected and undetected sources is $1.3\cdot10^{-22}$\,W/cm$^{2}$ and $1.1\cdot10^{-22}$\,W/cm$^{2}$, respectively. The median value for the hard band fluxes of the detected and undetected sources is $6.4\cdot10^{-22}$\,W/cm$^{2}$ and $6.7\cdot10^{-22}$\,W/cm$^{2}$, respectively. 

\citet{brandt2005} considered AGNs those whose soft band flux was larger than $5\cdot10^{-23}$\,W/cm$^{2}$. The median value of the fluxes in the soft band found for the VLBA detected sources is above this limit, which strengthens the assumption that they most likely contain an AGN. Nevertheless, the  
median value of the fluxes in the soft band found for the VLBA undetected sources is also above the limit suggested by \citet{brandt2005}, suggesting an AGN origin. We find a median value of the VLA integrated flux density for these sources of 120\,$\mu$Jy, making them amongst the faintest sources in the sample. At these flux densities sources are only detectable with the VLBA if the majority of their flux density comes from a compact component; this may explain the non-detection of these sources with the VLBA.


On the other hand, we find a relatively low number ($\sim$30\%) of X-ray counterparts for the VLBA detected sources. Since X-ray surveys are generally thought to be very efficient in finding AGNs (e.g. \citealt{mushotzky2004}), this deserves to be discussed. One possible explanation is that a certain type of AGNs is detected by radio surveys but not by X-ray surveys. Compton-thick AGNs with column densities of $N_{H} > 1.5 \cdot 10^{24} $cm$^{-2}$ (e.g. \citealt{treister2009}) are so heavily obscured that they remain undetected in X-ray surveys. However, they are not very common. \citet{lanzuisi2015} conducted a search of Compton-thick AGN in the XMM-COSMOS survey and found only ten. Another possibility is that we are looking at weakly-accreting AGN with lower accretion rates than the X-ray detected sources since it has been argued that hot-mode AGNs (radiatively inefficient) do not produce the X-ray characteristics of a typical AGN (e.g. \citealt{hardcastle2006}).

\begin{figure}
\centering
\includegraphics[scale=0.47]{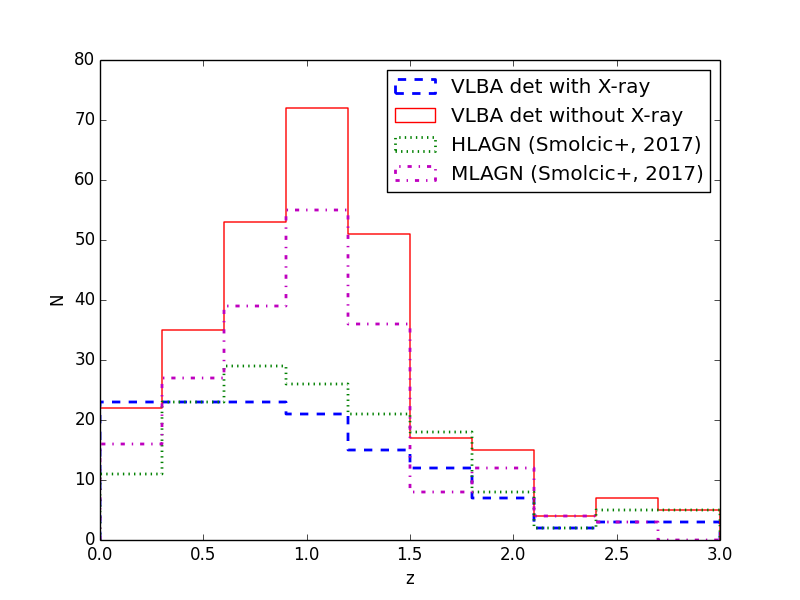}
\caption{\em {Redshift distribution of the VLBA detected sources with X-ray counterparts (dashed blue line) and without X-ray counterparts (solid red line). The VLBA detected sources classified either as HLAGN or as MLAGN by \citet{smolcic2017} are represented by the dotted green line (HLAGN) and the dash-dotted magenta line (MLAGN).}}
\label{fig:xraycount}
\end{figure}

\citet{smolcic2017a} presented the VLA-COSMOS 3\,GHz Large Project with which they observed the two square degree COSMOS field with the VLA at 3\,GHz. \citet{smolcic2017} studied the composition of the faint radio population selected from their VLA-COSMOS 3\,GHz Large Project and classified the radio sources as star forming galaxies or AGN. They further separate the AGN into moderate-to-high radiative luminosity AGN (HLAGN) and low-to-moderate radiative luminosity AGN (MLAGN) by using their multiwavelength properties. HLAGNs were selected by using a combination of X-ray, mid-infrared colour-colour and SED-fitting (see also \citealt{delvecchio2017}). The remaining sample was then classified as MLAGN, via excess of radio emission with respect to the star formation of the host galaxy, though completely silent in both X-rays and mid-infrared. We cross-matched our catalogue of VLBA detected sources with the sources classified as HLAGN or MLAGN in the catalogue of \citet{smolcic2017}. Figure~\ref{fig:xraycount} shows the redshift distribution of the VLBA detected sources with and without X-ray counterpart and the VLBA detected sources classified either as HLAGN or as MLAGN by \citet{smolcic2017}. We can see that the MLAGNs are mainly responsible for the bump of the VLBA detected sources without X-ray counterpart at z$<$1.5, possibly explaining why no X-ray counterpart is found for them. 

These results suggest that while X-ray surveys are highly efficient in selecting AGNs with high accretion rates (radiatively efficient), they may underestimate the number of AGNs when radiatively inefficient AGNs are considered since they may miss the AGNs with low accretion rates and therefore with low radiative luminosities.

\subsubsection{Radio-infrared correlation}

\begin{figure}
\centering
\includegraphics[scale=0.47]{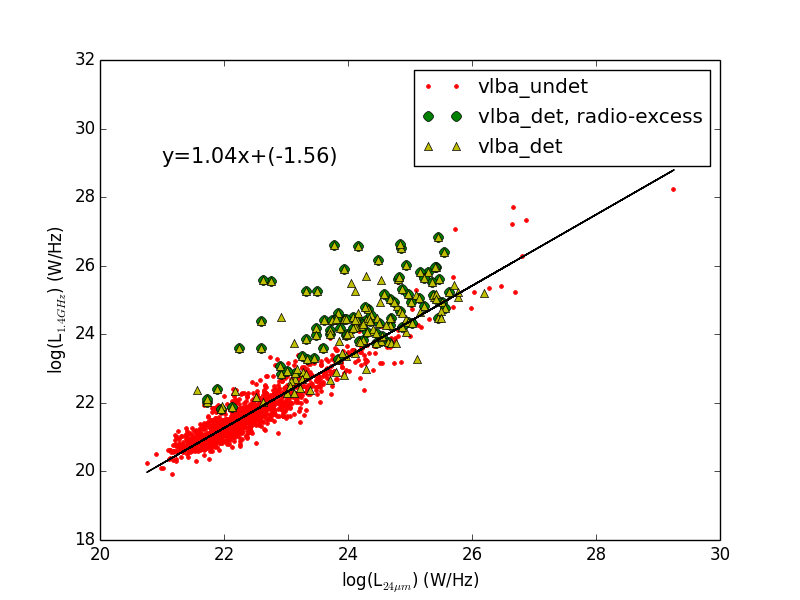}
\caption{\em {Radio(1.4\,GHz)\,-\,infrared(24\,$\mu$m) correlation of the sample, showing the VLBA detected sources classified as radio-excess sources by \citet{smolcic2017} (see text for details). The VLA flux densities have been used for all the plotted sources. The red dots represent the VLBA undetected sources. The yellow triangles represent the VLBA detected sources. The green circles represent the VLBA detected sources classified as radio-excess sources by \citet{smolcic2017}. The black line represents the linear regression fitting the VLBA undetected sources.}}
\label{fig:radioirsm}
\end{figure}

The radio-far-infrared correlation is a tight relation between the radio and far-infrared flux densities of galaxies \citep{condon1992}. A similar correlation has been demonstrated between the 24\,$\mu$m infrared flux density and the 20\,cm radio flux density \citep{appleton2004,boyle2007}. We have used the VLA flux densities to plot the radio-infrared correlation of our sample. We have assembled 24\,$\mu$m flux densities for 154 VLBA detected sources (33\%) and 1145 undetected sources (48\%) (see Section~\ref{sec:cat} for the references). Figure~\ref{fig:radioirsm} shows the position of the VLBA detected and undetected sources in the radio-infrared correlation. 


The radio-infrared correlation has usually been used to identify AGNs when there is a large radio excess, since the relation is thought to arise from star-forming activity. Nevertheless, \citet{roy1998} analysed the radio-far-infrared correlation using a sample of 149 Seyfert galaxies and radio-quiet quasars and found that many Seyferts displayed the same correlation between total radio and far-infrared emission as star-forming galaxies. Additionally, various studies have argued that many low-power AGN appear to obey this relation \citep{moric2010,obric2006}. Our results are in agreement with them since 53 VLBA detected sources (34\%) seem to follow the relation within 1$\sigma$. Because the majority of our sample ($\sim$70\%) consists of sub-mJy sources, this method of AGN identification might not be appropriate when the faint radio population is considered. 

\citet{smolcic2017} used the ratio of the VLA 1.4\,GHz radio luminosity and the star formation rate (SFR) in the host galaxy to classify the source as radio-excess source when the ratio showed an excess (see also \citealt{delvecchio2017}). The radio emission of the sources classified as radio-excess sources can be attributed to AGN activity. Figure~\ref{fig:radioirsm} also includes the VLBA detected sources classified as radio-excess sources by \citet{smolcic2017}. 96 of the 154 VLBA detected sources are also classified as radio-excess sources by \citet{smolcic2017}. Considering the 53 VLBA detected sources which seem to follow the relation, 20 of them are classified as radio-excess sources. These results suggest that the use of multiple techniques is important to identify all AGN. Amongst these, VLBA observations allow identification of AGN which are not picked up by other methods, and whose exclusion may otherwise lead to a biased view of the AGN population.

\begin{figure*}
\begin{subfigure}{\textwidth}
    \includegraphics[width=\textwidth, height=8.5cm]{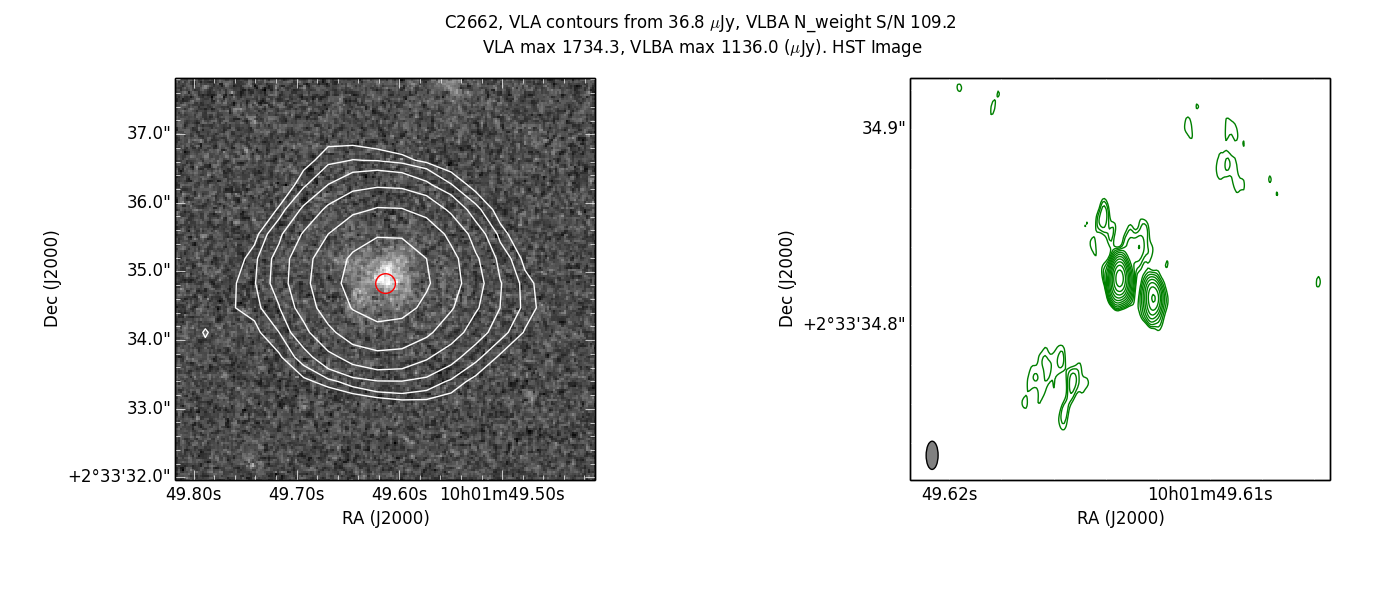}
\end{subfigure}
\begin{subfigure}{\textwidth}
    \includegraphics[width=\textwidth, height=8.5cm]{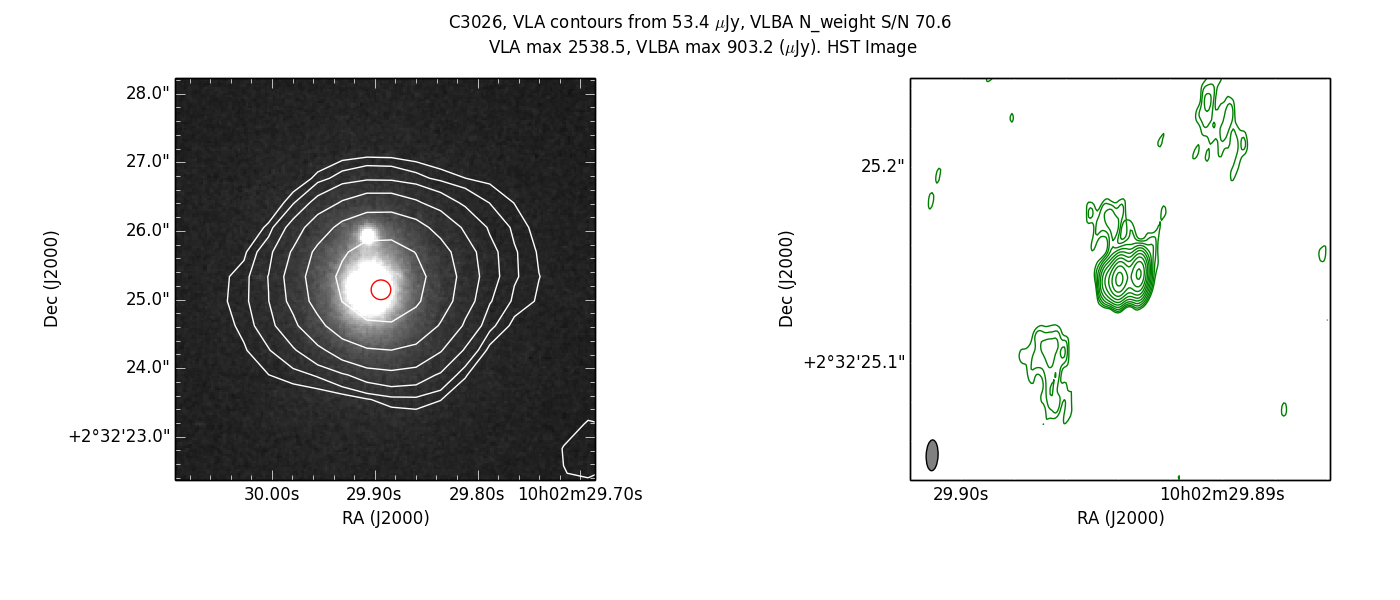}
\end{subfigure}
\caption{\em {Two VLBA detected sources, which exhibit a core with two-components. Left panel: The background greyscale image is the HST/Subaru image of the VLBA detection counterpart. The white contours represent the VLA contours of the source, starting at four times the rms noise level of the VLA image and increasing by a factor of two. The red circle represents the VLBA peak flux density position. Right panel: Green contours represent the VLBA detection contours, starting at three times the rms noise level of the naturally-weighted image and increasing by a factor of $\sqrt{2}$.}}
\label{fig:twoc}
\end{figure*}

\subsubsection{Morphology}

To study the optical morphology of the VLBA detected and undetected sources, we make use of the three classifications described by \citet{tasca2009}: 1) Early type; 2) Spirals; 3) Irregulars. We found counterparts for 327 detected sources (70\%) and 1547 undetected sources (65\%). 185 of the detected sources are classified as early type (57\%), 120 as spiral (37\%), and 22 as irregular (7\%). In the case of the undetected sources, 497 are classified as early type (32\%), 730 as spiral (47\%), and 320 as irregular (21\%). These results are in agreement with the findings from \citet{middelberg2013}, which suggested that the hosts of VLBI-detected sources are typically early-type or bulge-dominated galaxies.

We further separated the sample of the VLBA detected sources with morphological classification into low (z$<$1.5) and high (z$>$1.5) redshifts to compare the host optical morphologies. Out of the 283 VLBA detected sources at low redshift, 169 are classified as early type (60\%), 99 as spiral (35\%), and 15 as irregular (5\%). Out of the 44 VLBA detected sources at high redshift, 16 are classified as early type (36\%), 21 as spiral (48\%), and seven as irregular (16\%).  These results show that, unlike at low redshifts, at z$>$1.5 we found the major classification of host galaxies to be spiral (i.e., star-forming systems). This is in agreement with \citet{rees2016}, who investigated the host galaxy properties of a sample of radio-detected AGN to a redshift of z\,=\,2.25 and found that the majority of radio-detected AGN at z\,>\,1.5 are hosted by star-forming galaxies. This finding is interesting because these objects are unusual at low redshifts. One possible explanation is that at higher redshifts we might observe the transition between the starburst produced by a mayor merger and the triggering of the radio-loud AGN (e.g. \citealt{seymour2012}), which may lead to the later star formation quenching.



\subsection{VLBA-selected radio AGN and AGN selected by multiwavelength diagnostics}

\citet{delvecchio2017} analysed a sample of about 7900 radio sources in the COSMOS field observed with the VLA at 3\,GHz \citep{smolcic2017a} to explore the multiwavelength properties of AGN host-galaxies out to z\,$\lesssim$\,6. They used multiwavelength diagnostics to identify AGNs as described in Section~\ref{sec:xra}. To test the robustness of their method, they compared their source classification to other independent methods from the literature. In particular, they cross-matched their 3\,GHz VLA catalogue with our 1.4\,GHz VLBA catalogue and found that 91\% of the VLBA detected sources were classified as AGN by their method. 


The remaining 9\% of the VLBA detected source were misclassified as star-forming galaxies by \citet{delvecchio2017}, probably because they did not show AGN signatures in their multiwavelength properties (X-ray, mid-infrared and SED) and neither did show a significant radio-excess. We note that the threshold above which a radio excess was considered as significant by \citet{delvecchio2017} was rather conservative (about a factor of 5-6). By looking at those misclassified VLBA sources, they verified that most of them displayed systematically higher radio emission compared to the SFR in the host galaxy, but not significant enough to meet their above criterion (Delvecchio et al., priv. comm.).

\subsection{Two-compact-components VLBA sources}

Fig.~\ref{fig:twoc} shows two VLBA detected sources, which exhibit a core with two-components and extended emission not aligned with the two components. The appearance is similar to that found by \citet{rodriguez2006}, the only binary supermassive black hole (SMBH) widely accepted. The three possible scenarios explaining the features of these two VLBA detected sources are: i) one component is the core and the other is a knot in the jet, ii) the two components are the two lobes of a GPS radio source, or iii) the components are two nuclei involved in a supermassive black hole binary system.

The optical counterpart associated with the radio source C2662 is faint, with a V(AB) magnitude of 24.9. The photometric redshift is 1.4 and the separation between the two components is 176 pc. The optical image taken by the Hubble Space Telescope (HST) presents an irregular shape. 

The optical counterpart associated with the radio source C3026 has a V(AB) magnitude of 21.2. The photometric redshift is 0.43 and the separation between the two components is 63 pc. The image taken by the HST shows a normal elliptic shape. 

Spectral information is associated with the radio source C2662 from the Magellan Survey \citep{trump2007}. However, \citet{baldi2013} found that the object observed by Magellan was not the radio source C2662 (identified by them as ``Object 25''). No spectral information has been found for the candidate C3026.

A proposal to observe these two sources with the VLBA at a lower and a higher frequency to analyse the spectral energy distribution has been accepted in filler time. With these observations we expect to be able to discern between the three aforementioned hypothesis.

\subsection{VLBA-detected radio-quiet quasars}

The origin of the radio emission in radio-quiet quasars (RQQs) has been a matter of discussion for a long time. The two main scenarios ascribe it to either to the star forming activity of the host galaxy \citep{padovani2011, bonzini2013} or to the black hole activity of the AGN \citep{prandoni2010}. \citet{herreraruiz2016} reported for the first time on a sample of three RQQs where a lower limit of the radio emission coming from the AGN has been measured in the present project. The VLBA-measured radio flux densities are between 50\% and 75\% of the VLA flux densities, demonstrating that the radio emission of at least some RQQs is dominated by the black hole activity of the AGN. \citet{maini2016} have found similar results in a different sample, making this scenario more relevant.

\section{Conclusions}
\label{sec:con}

VLBI observations of 2865 radio sources were carried out with the VLBA at 1.4 GHz, obtaining milli-arcsecond resolution and a 1$\sigma$ rms noise level of 10\,$\mu$Jy in the central part of the field. The following points list the main outcomes of the project:

   \begin{enumerate}
      \item We have detected 468 sources with the VLBA. To date, this is the largest sample of VLBI detected sources in the sub-mJy regime. We have constructed a catalogue of the detected sources, the main objective of which is to be used as an AGN catalogue for future work in conjunction with complementary multi-wavelength data.
      \item On average, the VLBA recovered flux density of the detected sources is 60\% of the VLA flux density. This value represents a lower limit on the fraction of radio emission coming from the AGN. Additionally, we found that low flux density sources have a greater fraction of their radio luminosity in the core, suggesting that the faint radio population is indeed different from the brighter sources.
      \item The principal argument to consider the detected sources as AGNs is the high resolution provided by VLBI observations, which need sources with very high brightness temperatures to be detected. Moreover, after matching our observations with additional ancillary data, we can be mostly sure that our detections hold an AGN, mainly given their redshifts (we found a median photometric redshift of $\sim$1 for the VLBA detected sources). Nevertheless, no statement can be made for the undetected sources, since with the use of VLBI observations one can only positively identify AGNs, without implying that a non-detection is not an AGN. This follows because the AGN can be temporarily switched off or its emission can be below the detection threshold of the observations.
      
      \item The majority of the host galaxies of the VLBA detected sources are classified morphologically as early type galaxies, i.e., ellipticals and lenticulars, while the predominant classification for the VLBA undetected sources is spiral, in agreement with our expectations. Nevertheless, if we consider only the VLBA detected sources at z\,>\,1.5, the major classification of the host galaxies is spiral.

      \item The wide-field VLBI technique represents a powerful tool to distinguish radio source populations, what is very relevant for future observational projects with, for example, the SKA.
   \end{enumerate}

This project provides a valuable tool for the statistical analysis of the faint radio sky. Moreover, 25 hours of observations with the Green Bank Telescope (GBT) in addition to the VLBA have been awarded by the NRAO. We have observed one of the pointings of the COSMOS field with maximum sensitivity, to reach the faintest sources. These data together with the data from the present project will be matter of a future publication and will allow us to study the radio source counts from mJy to the tens of $\mu$Jy regime, revealing the AGN component of the faint radio population. 

\vspace{0.5cm}

\begin{acknowledgements}
      NHR acknowledges support from the Deutsche Forschungsgemeinschaft through project MI 1230/4-1. V.S. and I.D. acknowledge the European Union’s Seventh Framework programme under grant agreement 337595 (ERC Starting Grant, ``CoSMass''). PNB is grateful for support from STFC via grant ST/M001229/1. We wish to thank the anonymous referee for the helpful comments, which have improved this paper. This research made use of \texttt{Topcat} \citep{taylor2005}, available at \url{http://www.starlink.ac.uk/topcat/}. This research also made use of \texttt{APLpy}, an open-source plotting package for Python hosted at \url{http://aplpy.github.com}, and  \texttt{Astropy}, a community-developed core Python package for Astronomy \citep{astropy2013}. We wish to thank the staff of the VLBA who greatly supported the experimental observations in this project. The VLBA is operated by the Long Baseline Observatory, a facility of the National Science Foundation operated under cooperative agreement by Associated Universities, Inc.
\end{acknowledgements}



\setlength{\bibsep}{0pt plus 0.3ex}
\bibliographystyle{agsm}
\bibliography{references}

\end{document}